\documentclass[12pt]{article}

\usepackage[utf8]{inputenc}
\usepackage{amsmath,amssymb,amsthm,array,xcolor}
\allowdisplaybreaks

\usepackage[hyperfootnotes=false]{hyperref}
\hypersetup{
	pdftitle={Complex Menus in Discriminatory Pricing},
	pdfauthor={Patrick Lahr, Axel Niemeyer, Marek Pycia},
	pdfkeywords={},
	pdfcreator={LaTeX with hyperref package},
	pdfproducer={LaTeX},
	colorlinks=true,
	linkcolor=[rgb]{0,0,0.6},
	citecolor=[rgb]{0,0,0.6},
	urlcolor=[rgb]{0,0,0.6},
	pdfborder={0 0 0},
	pdfborderstyle={/S/U/W 0},
	pdfhighlight=/N,
	bookmarksnumbered=true,
	pdfdisplaydoctitle=true
}

\usepackage[pdftex]{graphicx}
\usepackage{fullpage}
\usepackage{enumitem}

\usepackage{subcaption}

\usepackage[backend=biber,style=authoryear-comp,maxbibnames=9,citetracker=true,maxcitenames=4,uniquename=false,uniquelist=false,natbib]{biblatex}

\usepackage[left=1in,right=1in,top=1in,bottom=1in]{geometry}
\usepackage{setspace}
\usepackage{titling}
\pretitle{\centering\large \bfseries}
\posttitle{\par\vskip 0.5em}
\preauthor{\centering\scshape\begin{tabular}[t]{c}}
	\postauthor{\end{tabular}\par}
\predate{\begin{center}}
	\postdate{\end{center}}

\usepackage{titlesec}
\titleformat{\section}{\centering\normalfont\scshape}{\thesection.}{1em}{}
\titleformat{\subsection}{\centering\normalfont\itshape}{\thesubsection.}{1em}{}

\titlespacing*{\subsection}{0em}{1em}{0.5em}

\usepackage{tikz}
\usepackage{pgfplots}
\pgfplotsset{compat=1.18}

\theoremstyle{plain}
\newtheorem{theorem}{Theorem}
\newtheorem{lemma}{Lemma}
\newtheorem{corollary}{Corollary}

\theoremstyle{definition}
\newtheorem{remark}{Remark}
\newtheorem{example}{Example}

\usepackage{etoolbox}
\expandafter\patchcmd\csname\string\proof\endcsname
{\normalparindent}{0pt }{}{}

\DeclareMathOperator*{\argmax}{arg\,max}
\DeclareMathOperator{\interior}{int}
\DeclareMathOperator{\relint}{relint}
\DeclareMathOperator{\menu}{menu}
\DeclareMathOperator{\REV}{REV}
\DeclareMathOperator{\mesh}{mesh}
\DeclareMathOperator{\Leb}{Leb}

\usepackage[nameinlink,noabbrev]{cleveref}
\crefname{theorem}{Theorem}{Theorems}
\crefname{corollary}{Corollary}{Corollaries}
\crefname{lemma}{Lemma}{Lemmas}
\crefname{appendix}{Appendix}{Appendices}
\crefname{figure}{Figure}{Figures}
\crefname{subsection}{Subsection}{Subsections}

\begin{document}
	\title{Complex Menus in Discriminatory Pricing\thanks{This work subsumes the paper ``Stochastic vs Deterministic Mechanisms in Multidimensional Screening'' of the third author (genericity and openness results from the 2000 DEEQA and 2006 PhD theses) and unpublished work of the first and second authors. For their comments, we thank Pete Caradonna, Piotr Dworczak, Glenn Ellison, Ernst Fehr, Thomas Graeber, Ilia Krasikov, Vijay Krishna, Jonathan Libgober, Noam Nisan, Pietro Ortoleva, Tom Palfrey, Luciano Pomatto, Jean-Charles Rochet, Eden Saig, Armin Schmutzler, Jakub Steiner, Jean Tirole, Yi Xin, Leeat Yariv, as well as audiences in Toulouse, MIT, Penn State, Princeton, Zurich, and at the Stony Brook International Conference on Game Theory. We are grateful for the hospitality of the Hausdorff Center for Mathematics (EXC-2047/2 – 390685813) and for the funding from the European Research Council (ERC) under EU Horizon 2020 grant agreement No 866376. AI tools were used in the preparation of this paper, including in simplifying and checking the proofs.}}

	\author{Patrick Lahr\thanks{ENS Paris-Saclay, Department of Economics. \textit{Email: \href{mailto:patrick.lahr@ens-paris-saclay.fr}{patrick.lahr@ens-paris-saclay.fr}}} \and Axel Niemeyer\thanks{Caltech, Division of the Humanities and Social Sciences. \textit{Email: \href{mailto:niemeyer@caltech.edu}{niemeyer@caltech.edu}}} \and Marek Pycia\thanks{University of Zurich, Department of Economics. \textit{Email: \href{mailto:marek.pycia@econ.uzh.ch}{marek.pycia@econ.uzh.ch}}}}

	\date{\today}
	\maketitle

\vspace{-1.3cm}

	\begin{abstract}

    \noindent
    Advances in technology decrease the costs of offering and choosing from large menus of varieties.
    As these complexity costs fall,
    how large are the menus offered by monopolists likely to become?
    For the multi-product or multi-attribute case, ubiquitous in practice,
    we show that optimal menus are generically infinite in the absence of complexity
    costs.
    Complexity costs can decrease the menu size, but it grows without bound as these costs decrease.
    These results have no analogue in the single-dimensional model and do not rely on non-linearities in production costs or preferences;
    the extensive product differentiation is caused by screening considerations alone.

    \end{abstract}

\section{Introduction}\label{sec:intro}

Technological progress shapes how firms design and price their product lines. The shift from mass production of standardized varieties to flexible manufacturing and mass customization in the late twentieth century is one example (e.g., \citet{milgrom1990economics}),
and ongoing advancements in information technology, including the recent developments in artificial intelligence (AI), are likely to be another. Information technology can radically decrease the costs of offering large menus of product varieties, including the cost of designing and creating the production capabilities for new varieties as well as the cost of administering them, setting optimal prices, and marketing them to consumers.
AI tools in particular can also reduce the buyer-side costs of evaluating large menus, costs that could otherwise induce buyers to disengage from the seller or retreat to other, simpler options.
We refer to these types of costs as complexity costs.\footnote{There are many recent examples of technology decreasing the costs of offering new product varieties
(\citet{babina2024artificial}; \citet{wsj2024mondelez}; \citet{reuters2026consumer}), and AI agents are on the cusp of being able to make shopping decisions for consumers (e.g., \citet{allouah2026aiagentbuyingevaluation}). In using the term complexity costs as an umbrella term,
we follow
usage in the operations management literature (e.g., \citet{Schaffer2008}).}

\begin{figure}[h]
    \centering
    \includegraphics[width=\textwidth]{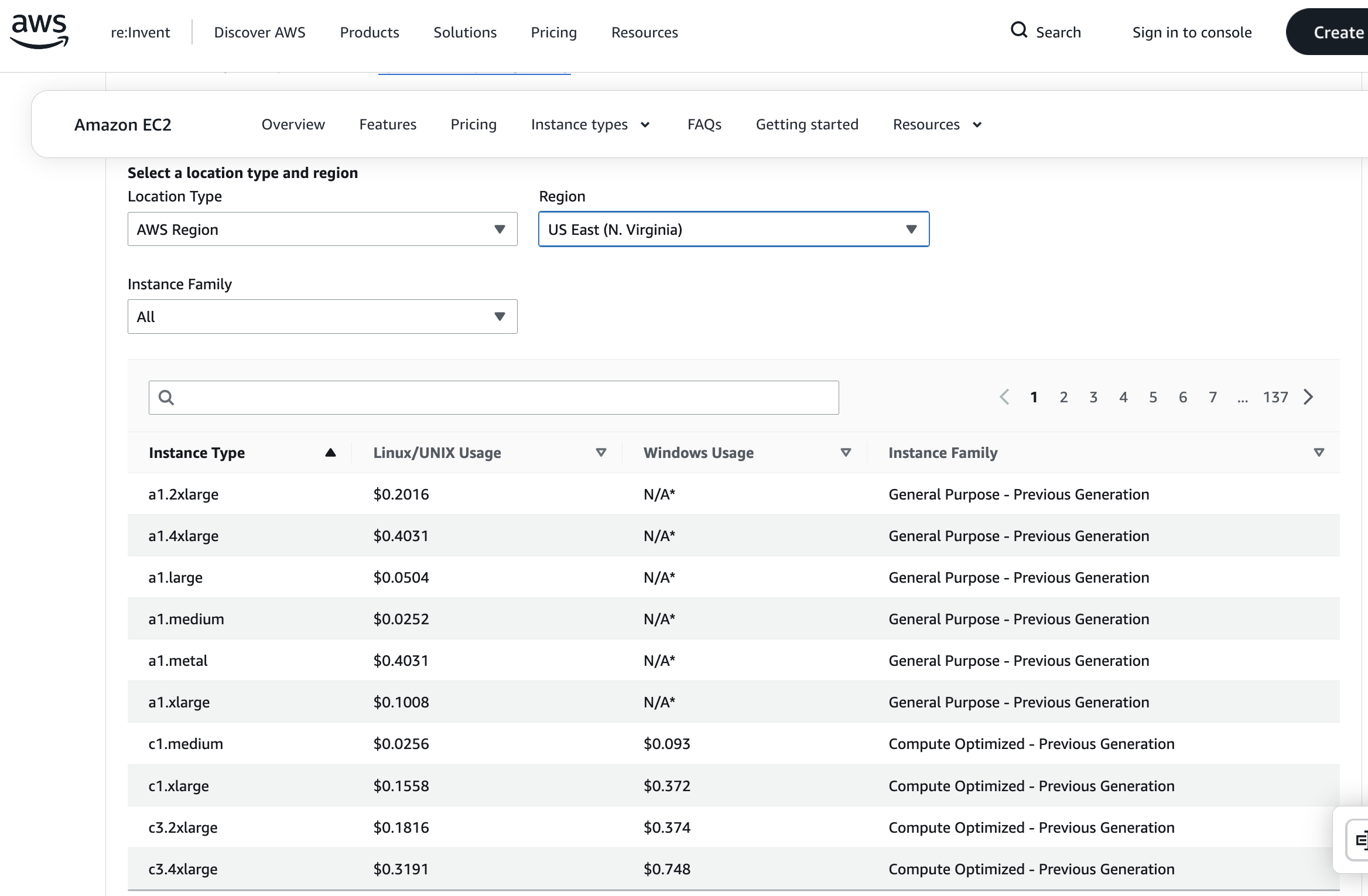}
    \caption{EC2 offers 1366 varieties in the region US East, here listed on 137 subpages with up to 10 varieties each.}
    \label{fig:EC2}
\end{figure}

Will we see substantially larger menus of varieties and contracts as complexity costs decrease? Or might optimal sales mechanisms continue to involve relatively small, simple menus even when offering many varieties is cheap?
While a larger menu allows the seller to tailor varieties more finely to heterogeneous tastes and to reach buyers with lower willingness to pay through cheaper, inferior versions, each new variety also cannibalizes the sales of existing ones.
How is this trade-off optimally resolved?
The classic results of \citet{stokey1979intertemporal}, \citet{myerson1981optimal}, and \citet{riley1983optimal} tell us that in single-dimensional screening models with linear costs and
valuations, selling just one variety is optimal.
Might an analogue of this insight continue to hold true in more realistic models that account for the multi-dimensional nature---with sellers marketing many products or products with multiple attributes---of most screening environments?
An illustrative example of what might happen instead is provided by the evolution of menu complexity observed in the cloud computing industry. Amazon Web Services’ EC2 (Elastic Compute Cloud) went from one product (instance) type in 2006 to over four hundred in 2021 and lists over 1300 instances in just one geographical region today,
as illustrated in Figure \ref{fig:EC2}.\footnote{See \citet{barr2021ec2} for the historical data.}

Our results suggest that cloud computing might well be a harbinger of things to come.
We focus our discussion on the canonical model of multi-dimensional monopoly pricing, in which product varieties are differentiated along several quality or quantity dimensions, and the seller's costs as well as buyers’ utility are linear in these dimensions (e.g., \citet{mcafee1988multi-dimensional}).
We study this model for several reasons.
First, it is the canonical benchmark in the multi-dimensional screening literature. Second, it captures well the setting of digital goods, which is where one might expect to see the largest impact of falling complexity costs in the near future.
Third, the linear model allows us also to isolate screening considerations in explaining complex menus.
Indeed, in the linear model, extensive product differentiation is inefficient and can arise solely from monopoly power.
Fourth, the model can be seen as a worst case for the need for differentiation, one in which we would least expect complexity, and, indeed, we show that our results extend beyond this canonical model.

The linear multi-dimensional screening model has been extensively studied in the absence of complexity costs, and we know that revenue-maximizing mechanisms are simple in some instances of the model and complex in others.
Indeed, an early version of the present paper constructed the first examples where menus of simple bundles in which each good (or attribute) is either included or excluded are not sufficient to maximize revenue, and, as we discuss below, more complex examples have been constructed since.%
\footnote{%
\label{fn:early-drafts}The present paper subsumes the first such examples from the 2000 DEEQA (MA) thesis of the third co-author; see \citet{pycia2006stochastic}, and also \citet{thanassoulis2004haggling} and \citet{manelli2006bundling}. For examples of more complex optimal mechanisms with infinitely many allocations, see \citet{daskalakis2017strong}, \citet{hart2019selling}, and \citet{lahr2024extreme}.}
It is also known that simple menus can perform poorly in some instances of the model.\footnote{See \citet{briest2015pricing}, \citet{hart2019selling}, and \citet{pycia2006stochastic}.}
However, these examples do not resolve the question of how prevalent the need for complex menus in these screening problems is.

In the canonical model without complexity costs, we show that revenue-maximizing menus for generic buyer populations are necessarily infinite.\footnote{{Genericity is here in the Baire category sense and obtains for various canonical spaces of type distributions. Menus are arbitrarily large for stronger notions of genericity.  \Cref{sec:result} provides details. }}
That is, the known instances where complex menus are needed for revenue-maximization are the norm rather than exceptions.
Accounting for complexity costs, we further show that large menus are generically optimal if complexity costs are small. In other words, if including an additional variety on the menu has only a small cost, then optimal menus will typically be large.
For seller-side costs, such as the costs of designing and marketing new product varieties, the optimal menu size increases monotonically as the costs decrease; for buyer-side costs, such as cognitive costs of choosing from the menu, convergence to large menus obtains as well but, interestingly, need not be monotonic.
Besides complexity costs, our results also extend to profit-maximization for a large class of production costs, in particular, when the cost of serving a buyer scales at least in proportion to the overall value delivered.\footnote{This property is satisfied whenever the seller can randomize allocations or the costs exhibit non-decreasing ray-average cost (RAC) in the sense of \citet{baumol1977proper}, which includes the standard specification of (strictly) convex costs (e.g., \citet{mussa1978monopoly}; \citet{rochet1998ironing}) and costs that are homogeneous of degree 1. }

The effects identified by our results may be most salient in the context of digital goods, such as cloud computing, software, or streaming media. This is because digital goods fit the linear monopoly model closely and because digital technologies lower complexity costs most directly for digital goods. Digital goods can often be differentiated along several dimensions, e.g., the number and type of processors, memory, or storage of cloud computing instances. Moreover, marginal costs are often negligible, as for software or streaming media, or scale at least in proportion to the value delivered, as seems reasonable for cloud computing.
Because digital goods are designed and most often delivered digitally, advances in digital technology translate directly into lower complexity costs. In particular, new varieties can be designed at low cost by removing or degrading product features (\citet{deneckere1996damaged}; \citet{varian2000versioning}), or by recombining or bundling features of existing products.\footnote{See also the literature in operations management on modular product architecture; e.g., \citet[Chapter 10]{ulrich2008product}.}
AI assistants may also help make large menus of products easier to navigate for consumers (\citet{allouah2026aiagentbuyingevaluation}).
The evolution of the EC2 menu described above is in line with these predictions.

    The complexity we establish is driven purely by screening considerations and is not a by-product of non-linearities or complementarities in the seller's cost or buyer's utility functions.
    Such non-linearities can drive menu complexity by themselves, and they are behind the complexity observed in optimal menus offered by a monopolist serving a single-dimensional population of buyers in \citet{mussa1978monopoly} and \citet{maskin1984monopoly} (or in the menu of contracts
    in the \citet{Baron-Myerson-1982} analysis of regulation).
    In their work, linear costs and preferences would have led to a simple reserve-price solution without any differentiation,  familiar from \citet{stokey1979intertemporal}, \citet{myerson1981optimal}, and \citet{riley1983optimal}.%
    \footnote{Equivalently, the quantity/quality would be sold via a single linear tariff. The single-variety expression of this claim is sometimes called the ``no-haggling'' result.}
    These same non-linearities also lead to product differentiation in multi-dimensional screening
    (\citet{wilson1993nonlinear}; \citet{armstrong1996multiproduct}; \citet{rochet1998ironing}).
    Importantly, in all of these models, separation is efficient, so product differentiation would arise in a competitive market as well.\footnote{\label{Mussa-Rosen-Axel}For example, in the model of \citet{mussa1978monopoly}, the monopolist offers a continuous product line. The offered products are the same as those in the efficient allocation, i.e., as those offered in a competitive market endowed with the monopolist's production technology, as long as some buyer types are excluded in the efficient allocation. See \citet{anderson2015product} for further discussion.}
    These models therefore cannot isolate screening considerations as the source of product differentiation.
    In our model, by contrast, fine differentiation is inefficient and arises solely as a consequence of monopoly power.
    Our results thus imply that the mere presence of asymmetric information generically leads to complex forms of price discrimination.

For generic type distributions, the optimal menu of price-variety pairs cannot be achieved by simple screening contracts such as two-part tariffs, despite these tariffs having an infinite menu. This is because the necessity of infinite menus for generic type distributions implies that optimal mechanisms must divide the market into infinitely many segments. With linear utility, however, buyers will never choose interior quantities from a two-part tariff (indifferences aside), leading to coarse market segmentations. In order to replicate optimal sales mechanisms with two-part tariffs for generic type distributions, the seller would have to offer an infinite menu of two-part tariffs with consumption caps.
Similarly, for a generic distribution, a menu of tapered tariffs with quantity discounts and consumption caps would also need to be infinite.
Our results thus imply that generic linear screening problems have highly non-linear solutions.

From this perspective, our results also provide an answer to a question known as Wilson’s conjecture:
why are observed pricing schemes relatively  simple?
Our results imply that the simplicity Wilson observed might have been driven by relatively large complexity costs.\footnote{\citet{wilson1993nonlinear} observed that, in practice, multidimensional tariffs price ``single aggregate measures that are easily appreciated by consumers'' and hypothesized that the costs of administration of more complex tariffs might be the underlying reason for this simplicity, a hypothesis we corroborate.
}
To the extent that technological progress lowers these costs, we would hence expect the premise of Wilson's conjecture to fail. Indeed, as noted above, the observed pricing of cloud computing is anything but simple.

We further note that our results do not hinge on menus being personalized to buyers' characteristics or to market segments, a personalization and segmentation that advances in information technology likely facilitate as well.\footnote{For recent analyses of personalized pricing in light of AI, see, e.g., \citet{shiller2020approximating} and \citet{dube2023personalized}. See also the theoretical literature following \citet{bergemann2015limits}.} Fine segmentation of the market here obtains already with pure second-degree price discrimination. Thus, even if personalized pricing is ruled out on legal grounds, our analysis suggests that technological advancements may still have a strong impact on how firms price discriminate.

Why is our result true in the multi-dimensional setting we study while its analogue fails in the one-dimensional case? Consider any optimal menu and a variety on the menu that would attract some of the currently excluded buyers with a small discount.
Imagine adding to the menu an inferior variety of half the quality at slightly less than half the price. Since utility is linear, this is effectively equivalent to randomizing between a small discount and an equally small markup on the original price. All else equal, randomizing a price cannot increase revenue. Hence, offering an inferior variety of half the quality cannot increase revenue either. This is why differentiation in the one-dimensional model never increases revenue; degrading products along a single dimension is all the seller can do. The revenue loss from offering the inferior variety is nevertheless of second order: it attracts previously excluded buyers but cannibalizes sales of the original variety, and these two effects roughly cancel. Now, with two or more attributes, the seller can perturb the inferior variety into a continuous array of similar varieties tailored to buyers' differences in taste for different attributes. The buyers of these new varieties form a region with curved boundaries in type space. No finite menu can replicate such curved boundaries, as it induces piecewise linear boundaries. Shifting a small mass of excluded buyers across these curved boundaries, toward the new varieties, leads to a first-order increase in revenue from the new varieties that small menus cannot replicate. Near any buyer type distribution, there are hence distributions for which larger menus yield strictly higher revenue. Moreover, if a distribution requires large menus, so do nearby distributions. Together, these two steps imply that optimal menus are generically infinite. \Cref{sec:result} makes this precise.

\paragraph{Related Literature.}
It is known in the multidimensional screening literature that revenue-maximizing mechanisms may need to use fractional allocations (interior qualities, probabilities, etc.); an  early draft of the present paper provided the first example of this phenomenon and showed that it obtains for generic type distributions (see \Cref{fn:early-drafts}).
But fractional allocations per se do not make a mechanism complex. For instance, a menu consisting of a basic version and a premium version of a product can hardly be considered complex just because some product features are graded in the basic version. We therefore follow \citet{hart2019selling} and \citet{babaioff2017menu} in measuring the complexity of a mechanism by its menu size, the number of distinct allocations across all possible types; as discussed above, our results are also valid for measures of complexity based on the number of two-part tariffs the seller would need to offer to the buyers.\footnote{Other notions of complexity are studied by \citet{daskalakis2014complexity}, \citet{dughmi2014sampling}, and \citet{morgenstern2015pseudo}; see also \citet{li2017obviously} and \citet{pycia2023theory}.}

A striking finding in the subsequent literature is that there exist type distributions for which infinite menus are necessary for optimality. \citet{daskalakis2017strong} give a two-good example with i.i.d.\ values with simple, linear densities in which every optimal mechanism has an infinite menu. \citet{hart2019selling} construct sequences of type distributions along which menus up to any given finite size obtain only a vanishing fraction of maximal revenue. More generally, \citet{lahr2024extreme}, building on \citet{manelli2007multi-dimensional}, show that virtually every incentive-compatible mechanism can be rationalized as uniquely optimal for some type distribution.

None of these papers address how frequently optimal menus are complex. Our contribution is to settle this question: generic type distributions necessitate complex menus.

Another burgeoning strand of the multidimensional screening literature identifies environments in which simple mechanisms are optimal (\citet{mcafee1989multiproduct}; \citet{manelli2006bundling}; \citet{fang2006bundle}; \citet{pavlov2011optimal}; \citet{daskalakis2017strong}; \citet{tang2017optimal}; \citet{menicucci2015optimality}; \citet{bergemann2021optimality}; \citet{haghpanah2021pure}; \citet{ghili2023characterization}; \citet{yang2023nested}; \citet{haberman2025multi-dimensional}; \citet{pernoud2025bundling}).
Within the present model, such simplicity results must rest on special, non-generic conditions. Of course, this does not diminish the value of these results. On the contrary, they identify those important environments in which the model departs from its tendency towards complexity. However, our genericity analysis suggests that if simple mechanisms are observed in practice, then other, unmodeled factors are likely at play. One important such factor considered in this paper is complexity costs, discussed in \Cref{sec:discussion}.

A strand of the literature has investigated when the classical model of vertical differentiation with one-dimensional types predicts price discrimination at all (\citet{deneckere1996damaged}; \citet{johnson2003multiproduct}; \citet{mcafee2007pricing}; \citet{bhargava2008research}; \citet{anderson2009price}).
This question is non-trivial because, with linear utility and constant marginal costs, it is optimal for the seller to offer only the best possible product at a fixed price (\citet{stokey1979intertemporal}; \citet{myerson1981optimal}; \citet{riley1983optimal}).
By contrast, with cost convexities as in \citet{mussa1978monopoly} or decreasing marginal utility as in \citet{maskin1984monopoly}, it is efficient to offer differentiated products, i.e., product differentiation is a feature even of a perfectly competitive market (see  \Cref{Mussa-Rosen-Axel}).
In order for there to be no product differentiation in the efficient solution but differentiation in the monopoly solution, specific assumptions on consumer utility, in particular departures even from multiplicatively separable utility, are needed.
In two or more dimensions, by contrast, our result shows that even linear utility---the most widely used utility specification---generically implies a strong motive for the seller to price-discriminate despite it being inefficient.

By showing that complexity costs or other frictions are generically needed to rationalize simple screening menus, we also contribute to
various strands of
the literature
on complexity costs.
On the seller side, setup costs are a classical explanation for limited product variety offered by multi-product monopolies with horizontally differentiated product lines
(see, e.g., the survey by \citet{lancaster1990economics}).
By contrast, we focus on the case of vertical differentiation, where all buyers agree on the best variety and where it would be efficient to offer only a small number of varieties, and identify the optimal number of varieties as setup costs and other complexity costs vanish.\footnote{See also \citet{netessine2007product} and \citet{wong2014choice}, who study the optimal number of varieties in single-dimensional models with vertical differentiation as in \citet{mussa1978monopoly}. }
On the buyer side, the cognitive costs we consider can reflect the effort of examining a menu, discovering one's valuations, or identifying the utility-maximizing option.%
\footnote{%
Such costs are usually
microfounded, e.g.,  via the cost of inspecting  options in turn (\citet{Weitzman-ECTA-1979}), overcoming information overload (e.g.,
\citet{Bertrand-Dean-Mullainathan-QJE-2010}),
or attention (e.g., \citet{Geanakoplos-Milgrom-attention-1991}, \citet{matejka2015rational}, and \citet{Caplin-Dean-Leahy-REStud-2019}).
An optimal response to these costs might be to balance the cost of
analysis with gains from choosing the better option. In such cases, the buyer's optimal choice might be random. While we focus on the case where, after incurring the cost, the buyer can choose the best option, our analysis can be extended to random choice provided the random choice converges to the best option as the costs vanish.
}
A rich literature documents the direct impact of large menus on choice behavior, including effects such as a retreat to the outside option (\citet{iyengar2000choice}), the choice of a default (\citet{De-Lara-Dean-2025}), or the choice of the middle (compromise) option (\citet{Simonson-Tversky-JMR-1992}).
The indirect impact of a specific type of complexity cost has been studied by, e.g.,
\citet{Gabaix-2026}, who shows that the dread associated with making an incorrect choice may simplify prices as well as contracts and tax systems;
our corollary on the impact of buyer-side complexity costs  on pricing allows a wide range of cost specifications.

In the following, we first study the canonical model without complexity costs (\Cref{sec:definitions,sec:result}) and then develop our results on complexity costs and other frictions, as well as an extension to general cost structures  (\Cref{sec:discussion}).

\section{Model}
\label{sec:definitions}
We begin by studying a model without costs---or, equivalently, by studying revenue maximization---and incorporate production and complexity costs in \Cref{sec:discussion}.

    A seller sets prices for allocations $q\in Q=[0,1]^2$. Allocations may represent, for instance, products that are vertically differentiated along several dimensions, quantities of different products, or probabilities of obtaining different products.
    The restriction to a two-dimensional problem is for ease of exposition but inessential for the results (see \Cref{higher-dimensions}).

    These allocations are desired by a population of potential buyers. Buyers are indexed by their type $\theta=(\theta_1,\theta_2)\in\Theta=[0,1]^2$, representing their marginal allocation values: if buyer type $\theta=(\theta_1,\theta_2)$ obtains allocation $q=(q_1,q_2)$ and pays $p$ to the seller, then the buyer's utility is $q\cdot\theta-p$. Buyers can refuse to trade and obtain a utility of zero.
	The population distribution of buyer types is given by a Borel probability measure $\mu$ on $\Theta$. The seller knows the distribution $\mu$ but cannot distinguish individual buyer types.

	By the revelation principle, the seller can focus on incentive compatible (IC) and individually rational (IR) direct mechanisms (henceforth, mechanisms). A mechanism $M=(x,t)$ consists of Borel measurable allocation and payment rules $x:\Theta\to Q$ and $t:\Theta\to\mathbb{R}_+$ assigning to each buyer type $\theta$ an allocation $(q_1,q_2)$ and a payment $p$, respectively.\footnote{Non-negative payments are without loss of generality for revenue maximization.} IC means that buyers are incentivized to reveal their type truthfully, and IR means that buyers are incentivized to participate:
	\begin{alignat}{2}
		\tag{IC}
		\label{eq:IC}
		x(\theta)\cdot\theta-t(\theta)&\geq x(\theta')\cdot\theta-t(\theta') \qquad&&\forall\theta,\theta'\in\Theta\\
		\tag{IR}
		\label{eq:IR}
		x(\theta)\cdot\theta-t(\theta)&\geq 0 \quad&&\forall \theta\in\Theta.
	\end{alignat}
	We write $x_M$ and $t_M$ for the allocation and payment rules of $M$ whenever we need to index the mechanism. Let $\mathcal{M}$ denote the set of all IC and IR mechanisms.

	By the taxation principle, the seller can equivalently offer a menu of allocation-price pairs $(q;p)$ and let buyers choose their favorite option. The menu associated with mechanism $M=(x,t)$ is given by
	$$
	\menu(M)=\{(x(\theta);t(\theta))\}_{\theta\in\Theta}\cup\{(\mathbf{0};0)\}.
	$$
	IC and IR mean that $(x(\theta);t(\theta))$ is one of $\theta$'s favorite options in $\menu(M)$.\footnote{Following \citet{babaioff2017menu}, we require without loss of generality that any  menu contains the outside option $(\mathbf{0};0)$. This is convenient for some arguments but otherwise inessential.} We refer to the cardinality $\lvert\menu(M)\rvert$ as the \textbf{menu size} of $M$. The menu size of $M$ is the smallest number of options needed to implement $M$ or, equivalently, the number of bunches or market segments induced by $M$ (modulo the outside option).

	The seller's revenue from mechanism $M$ for type distribution $\mu$ is given by
	$$
		\REV(M;\mu)=\int_\Theta t_M(\theta)\,d\mu(\theta).
	$$
    We also write $\REV(M;f)$ if $\mu$ has a density $f$ with respect to Lebesgue measure.
	A mechanism is \textbf{optimal} if it maximizes $\REV(M;\mu)$ across all $M\in\mathcal{M}$. An optimal mechanism exists (e.g., \citet{carlier2020existence}).

	\section{The Genericity of Complex Menus}
	\label{sec:result}

    We now show that revenue maximization requires complex menus for generic type distributions, and we sketch the proof.

    Key to the formulation of our result are the following notions of genericity: (i)
    a set is \textbf{generic} if it contains a dense set that is also a countable intersection of open sets; (ii) a set is \textbf{strongly generic} if it contains an open and dense set.\footnote{That is, ``generic'' is here in the Baire category sense. We use the term ``strongly generic''
    in order to distinguish the weaker from the stronger property.
    For earlier uses of this terminology, see, e.g., \citet{Broer-Takens-2010}.
    As our proof shows, the strongly generic claim of our main result, \Cref{thm:master}, actually holds on an open and dense set, so the property of large menu size is stable under perturbations.
    }
    See, e.g., Section 8.2 in \citet{mas1985theory} for further discussion on genericity.
    Since our genericity notions are topological, we must specify a topology or metric for the space of type distributions.

    Unless mentioned otherwise, our results hold for each of the following spaces of type distributions:
	\begin{enumerate}[nosep]
		\item the space $\mathcal{P}$ of Borel probability measures on the type space $\Theta$ equipped with the weak topology (relative to the canonical duality with bounded, continuous functions), metrized here by the Wasserstein or Prokhorov distance;
		\item the space $\bar{\mathcal{F}}=\{f\in L^1(\Theta)\mid f\geq 0,\, \int f=1\}$ of probability densities equipped with the $L^1$-norm $\|f\|_1=\int |f|$ (equivalently, Lebesgue absolutely continuous probability measures equipped with the total-variation distance);
		\item the space $\mathcal{F}\subset\bar{\mathcal{F}}$ of continuous probability densities
        equipped with the sup-norm $\|f\|_\infty=\sup_{\theta\in\Theta} |f(\theta)|$.
	\end{enumerate}
    The spaces are progressively smaller whereas the topologies are progressively stronger.

	\begin{theorem}
		\label{thm:master}
		For every $k\in\mathbb{N}$, the set of type distributions for which every optimal mechanism has menu size greater than $k$
        is strongly generic.
        In particular, for generic type distributions, every optimal mechanism has infinite menu size.
	\end{theorem}

    \begin{proof}
        To establish the first claim, we show that the set is open and dense in each of the three spaces; see \Cref{subsec:openness} for the former and \Cref{subsec:denseness} for the latter.
        Intersecting over all $k\in\mathbb{N}$ and applying the Baire Category Theorem yields the second claim.
    \end{proof}

We first comment on the theorem and its corollaries and then discuss the proof.

    \Cref{thm:master} predicts that revenue-maximizing menus are generically complex when a monopolistic seller can differentiate products along multiple dimensions without any frictions. We reiterate that this prediction is driven purely by informational asymmetries between seller and buyer; non-linearities in production costs or preferences are not needed to explain product differentiation as a result of price discrimination. Indeed, since differentiation is inefficient in our model, it is solely a consequence of monopoly power and would not arise in a competitive market (unlike in the classical models of \citet{mussa1978monopoly}, \citet{maskin1984monopoly}, \citet{armstrong1996multiproduct}, and \citet{rochet1998ironing}).
    Although our result has direct implications for the pricing and product line design of digital goods, where marginal production costs are often negligible, it also extends to a large class of (not necessarily convex) cost functions. Since complex menus are rarely seen in practice, especially in business-to-consumer contexts, the theory suggests that other, hitherto unmodeled factors such as complexity costs are likely at play in rationalizing simple menus. We elaborate on all of these points in much more detail in \Cref{sec:discussion}.

    We emphasize that \Cref{thm:master} quantifies over all optimal mechanisms. This is important because the optimal mechanism is generally not unique. In particular, there exist spurious optimal mechanisms with infinite menu size that arise solely due to tie-breaking by indifferent types.\footnote{To illustrate this, suppose the seller offers $(1,1)$ at $p=1$ and the outside option.
    It can be shown that this menu is optimal for some absolutely continuous type distribution $\mu$.
    If we parametrize the indifference line $\theta_1+\theta_2=1$ by $\ell(s)=(s,1-s)$, then the mechanism that allocates option $(s,s;sp)$ to type $\ell(s)$ and leaves the allocation of all types off the indifference line unchanged is also optimal for $\mu$ and has infinite menu size. Similar constructions can be carried out for any mechanism of finite menu size.}
    To reason about uniqueness, we therefore restrict attention to seller-favorable mechanisms (cf. \citet{hart2015maximal}). A mechanism is \textbf{seller-favorable} if, among all mechanisms that induce the same utility for every buyer type, it maximizes the payment of every type and, if a tie remains for some types, gives such types the lexicographically largest allocation.

    \begin{corollary}
		\label{cor:uniqueness}
		The set of densities for which the optimal seller-favorable mechanism is unique and has infinite menu size is
         generic
        in each of $\bar{\mathcal{F}}$ and $\mathcal{F}$.
	\end{corollary}

    \begin{proof}
        See \Cref{subsec:uniqueness}.
    \end{proof}

The above theorem and corollary have direct implications for a screening problem in which buyers are asked to choose a two-part tariff and then select an allocation and pay the price determined by the tariff's pricing function.
If each tariff allows the buyer to choose any allocation from $[0,1]^2$ then
the linearity of buyers' values implies that each buyer type would select one of the extreme points $(0,0),(0,1),(1,0),(1,1)$ (if indifferences are broken in the seller's favor). Thus, our theorem implies that menus of such two-part tariffs generically cannot be revenue maximizing.
In practice, some tariffs restrict the domain of allocations the buyer has access to. This is the case, e.g., in the Swiss health insurance system where the tariff chosen not only specifies the premium and copays but also the availability of some services and medical doctors. Similarly, the ``Tarif Bleu'' offered by the French electric utility company EDF combines a fixed subscription charge and per-kWh prices with a subscribed-power limit.
To allow for such restrictions, we define a \textbf{two-part tariff} by its consumption caps $\underline{q}_1,\overline{q}_1 , \underline{q}_2,\overline{q}_2 \in [0,1]$ and an affine pricing function $P:Q_P = [\underline{q}_1,\overline{q}_1 ]\times [\underline{q}_2,\overline{q}_2 ]\to \mathbb{R}$.
Having selected such a tariff, the buyer chooses an allocation $q\in Q_P$ and pays $P(q)$.
The linearity of buyers' preferences then again implies that each buyer would select from such a tariff one of the extreme points of the rectangle $Q_P$.
Hence \Cref{thm:master} implies:
\begin{corollary}
    	\label{cor:tariffs}
        For every $k\in\mathbb{N}$, the set of type distributions for which every optimal menu of two-part tariffs offers at least $k$ tariffs
        is strongly generic.
        In particular, for generic type distributions, every optimal menu of two-part tariffs has infinite size.
        Without consumption caps, the set of type distributions for which no menu of two-part tariffs is
     optimal is strongly generic.
\end{corollary}

An analogue of this corollary, with the same argument, obtains for \textbf{tapered tariffs} (or tariffs with quantity discounts), defined by their consumption caps $\underline{q}_1,\overline{q}_1 , \underline{q}_2,\overline{q}_2 \in [0,1]$ and a pricing function $P:Q_P = [\underline{q}_1,\overline{q}_1 ]\times [\underline{q}_2,\overline{q}_2 ]\to \mathbb{R}$ that is concave in each of the arguments.

    \subsection*{Proof Sketch}

 We now give a detailed sketch of the proof of \Cref{thm:master}. Readers may skip ahead to \Cref{sec:discussion} without loss of continuity.

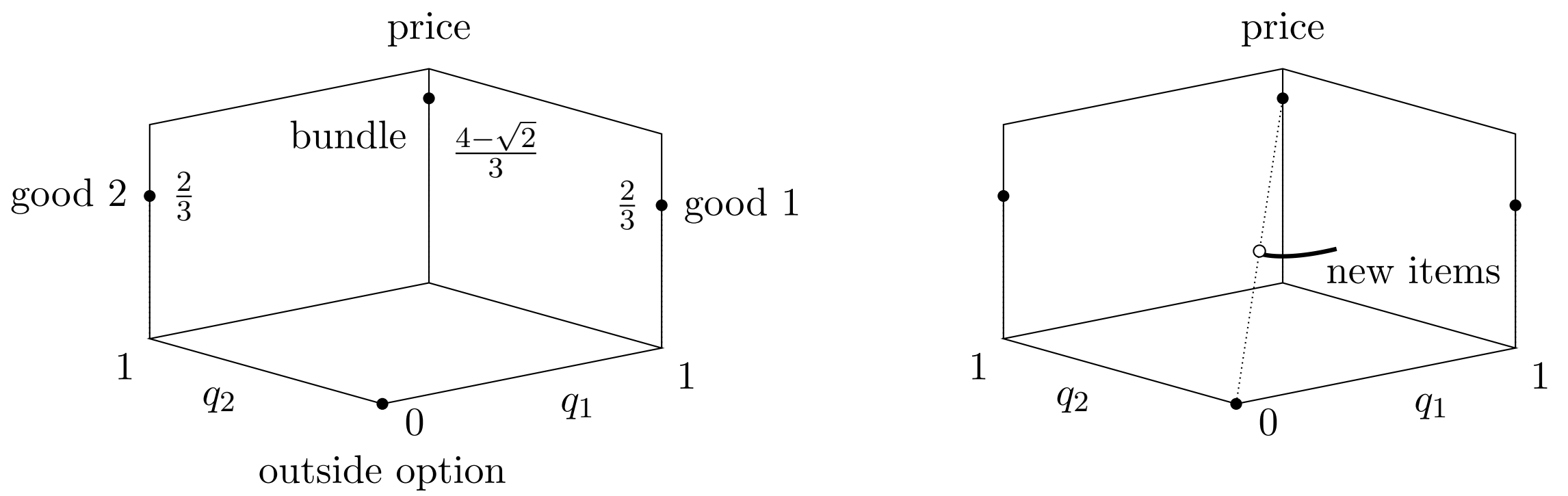
\begin{figure}[t]\centering\singlespacing
\tikzset{declare function={eps=1/90; p=(4-sqrt(2))/3; sleft=-0.350648; sright=0.442748;%
    phi(\s)=1+2.5*(\s+.8)^4; dphi(\s)=10*(\s+.8)^3;
    qnew(\s)=1/2+eps*dphi(\s)/sqrt(2); pnew(\s)=p/2+eps*(\s*dphi(\s)-phi(\s));}}
\newcommand\cube[1]{\begin{tikzpicture}[x={(3cm,.6cm)},y={(-2.5cm,.7cm)},z={(0cm,2.3cm)},scale=.9,font=\small]
  \draw (0,0,0)--(1,0,0)--(1,1,0)--(0,1,0)--cycle (1,0,0)--(1,0,1)--(1,1,1)--(0,1,1)--(0,1,0) (1,1,0)--(1,1,1);
  \path (0,0,0) node[anchor=west,xshift=2pt,yshift=-5pt] {$0$} (1,0,0) node[below right] {$1$} (0,1,0) node[below left] {$1$}
    (1,1,1) node[above=2pt] {price} (.7,0,0) node[below=4pt] {$q_1$} (0,.7,0) node[below=4pt] {$q_2$};
  \foreach \x/\y/\z in {1/0/{2/3}, 0/1/{2/3}, 1/1/{p}}{\draw[densely dotted,thin] (\x,\y,0)--(\x,\y,\z); \fill (\x,\y,\z) circle[radius=1.8pt];}
  \fill (0,0,0) circle[radius=1.8pt] node[below=10pt] {\phantom{outside option}};%
  #1\end{tikzpicture}}
\begin{subfigure}[b]{.5\textwidth}\centering
  \cube{\path (0,0,0) node[below=10pt] {outside option}
    (1,0,{2/3}) node[right=2pt] {good 1} node[left=2pt] {$\tfrac23$} (0,1,{2/3}) node[left=2pt] {good 2} node[right=2pt] {$\tfrac23$}
    (1,1,{p}) node[below left=2pt] {bundle} node[below right=2pt] {$\tfrac{4-\sqrt2}{3}$};}
  \caption{$M$}\label{subfig:menu-M}
\end{subfigure}%
\begin{subfigure}[b]{.5\textwidth}\centering
  \cube{%
    \draw[densely dotted,thin] (0,0,0)--(1,1,{p});
    \draw[very thick] plot[domain=sleft:sright,variable=\s,samples=100] ({qnew(\s)},{1-qnew(\s)},{pnew(\s)});
    \draw[fill=white] (.5,.5,{p/2}) circle[radius=1.8pt];
    \node[below right=-2pt,xshift=-5pt] at ({qnew(sright)},{1-qnew(sright)},{pnew(sright)}) {new items};}
  \caption{$M^\varepsilon$}\label{subfig:menu-Meps}
\end{subfigure}
\caption{The menu of the mixed bundling mechanism $M$ and its perturbation $M^\varepsilon$.}
\label{fig:menus}
\end{figure}

      A classical example due to \citet{manelli2006bundling} shows that an optimal mechanism $M$ for types that are distributed uniformly on the unit square has menu size four: $(0,1)$ and $(1,0)$ at price $2/3$, $(1,1)$ at price $p=(4-\sqrt2)/3$, and the outside option $(0,0)$ at price $0$. The menu is depicted in \Cref{subfig:menu-M}. The mechanism $M$ represents mixed bundling of two goods with no further quality differentiation.

    We now illustrate how to perturb the uniform distribution such that mixed bundling is no longer optimal.
    While we focus on this example, the extension to the non-uniform case and any optimal mechanism with finite menu size is straightforward, as we explain later.
    In fact, our construction shows that no mechanism of menu size $\leq k$ is optimal, for arbitrary $k\in\mathbb{N}$.
    This construction is the more involved part of the proof. Openness, by contrast, is relatively straightforward. It holds because the gap between optimal revenue and the best revenue from at most $k$ items is continuous; see the  proof.

    We construct below, for all sufficiently small $\varepsilon>0$, a mechanism $M^\varepsilon$ with infinite menu size and a signed density $g^\varepsilon$ with $\int g^\varepsilon=0$ and $|g^\varepsilon|\leq1$. The signed density represents a mass transport. The idea is that $M^\varepsilon$ loses revenue on the order of $\varepsilon^2$ under the uniform density $f$ but has a uniform advantage under $g^\varepsilon$ on the order of $\varepsilon^{5/4}$ against every competing mechanism $M'$ with at most $k$ items. Consequently, for $f^\varepsilon=f+\sqrt{\varepsilon}g^\varepsilon$, $\REV(M^\varepsilon;f^\varepsilon)>\REV(M';f^\varepsilon)$ for all sufficiently small $\varepsilon>0$. Moreover, $f^\varepsilon$ is a well-defined density with $f^\varepsilon\to f$. $M^\varepsilon$ need not be optimal for $f^\varepsilon$.

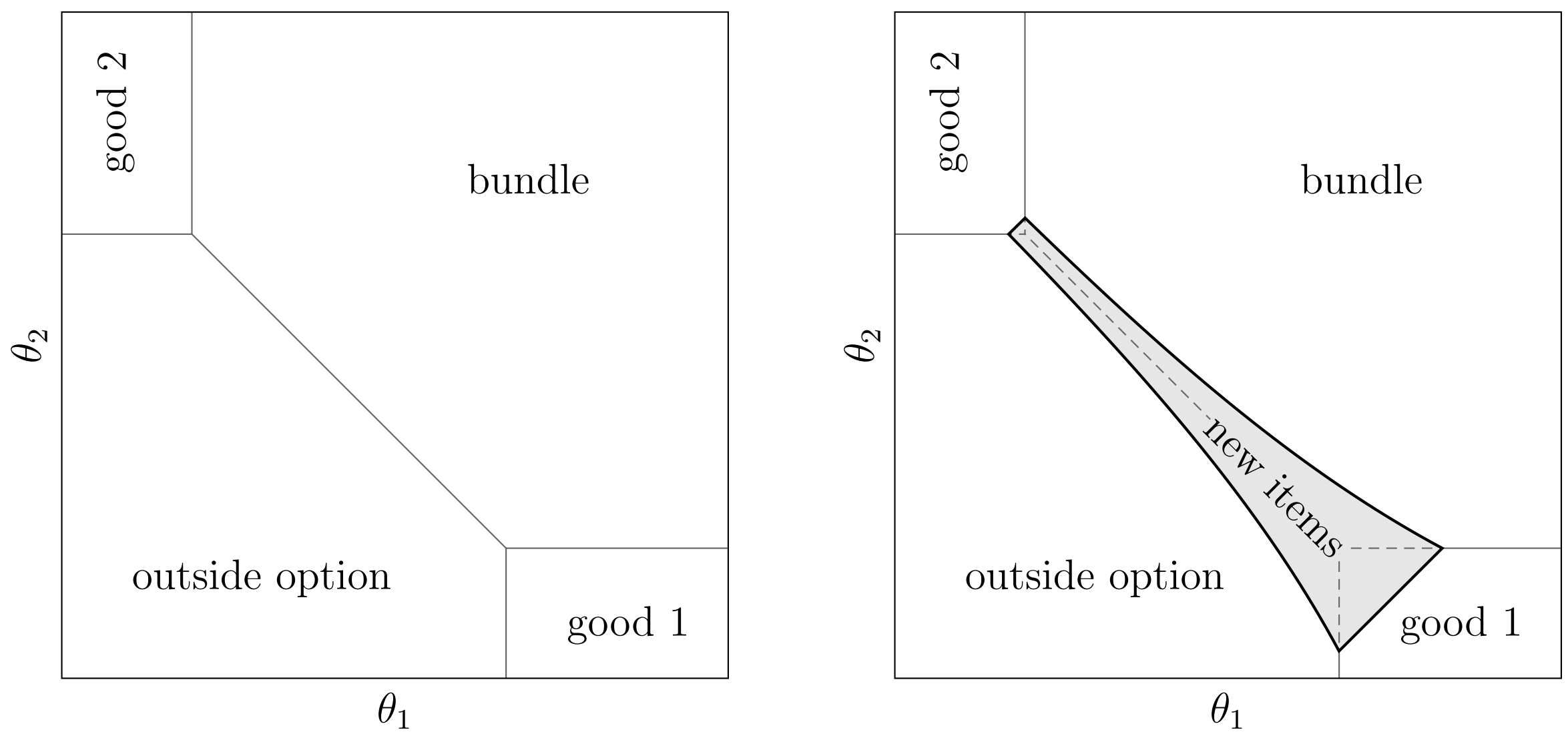
\begin{figure}[t]\centering\singlespacing
\tikzset{declare function={eps=1/90; p=(4-sqrt(2))/3; single=2/3; excess=p-single; sleft=-0.350648; sright=0.442748;%
    phi(\s)=1+2.5*(\s+.8)^4; c(\s)=2*eps*phi(\s);
    tx(\s,\l)=(p+\l+sqrt(2)*\s)/2; ty(\s,\l)=(p+\l-sqrt(2)*\s)/2;},
  every plot/.style={variable=\s,samples=100}}
\newcommand\pt[2]{axis cs:{tx(#1,#2)},{ty(#1,#2)}}%
\newcommand\band[1]{\path[#1] plot[domain=sleft:sright] (\pt{\s}{-c(\s)}) -- plot[domain=sright:sleft] (\pt{\s}{c(\s)}) -- cycle;}
\newcommand\oldcells[1]{\draw[#1] (axis cs:{single},0)--(axis cs:{single},{excess})--(axis cs:1,{excess})
  (axis cs:0,{single})--(axis cs:{excess},{single})--(axis cs:{excess},1) (axis cs:{single},{excess})--(axis cs:{excess},{single});}
\newcommand\typepanel[1]{\begin{tikzpicture}\begin{axis}[scale only axis,width=.8\linewidth,height=.8\linewidth,
    axis lines=box,axis on top,xmin=0,xmax=1,ymin=0,ymax=1,xtick=\empty,ytick=\empty,
    xlabel={$\theta_1$},ylabel={$\theta_2$},label style={font=\normalsize}]
  #1
  \node at (axis cs:.3,.15) {outside option}; \node at (axis cs:.85,.08) {good 1};
  \node[rotate=90] at (axis cs:.08,.85) {good 2}; \node at (axis cs:.7,.75) {bundle};
  \end{axis}\end{tikzpicture}}
\begin{subfigure}[b]{.5\textwidth}\centering
  \typepanel{\oldcells{black!60,thin} (axis cs:{single},{excess})--(axis cs:{excess},{single});}
  \caption{$M$}\label{subfig:types-M}
\end{subfigure}%
\begin{subfigure}[b]{.5\textwidth}\centering
  \typepanel{\oldcells{black!60,thin}\band{fill=black!10}
    \begin{scope}\band{clip}\oldcells{black!60,thin,densely dashed}\end{scope}
    \node[fill=black!10,inner sep=.5pt,rotate=-45] at (\pt{.2}{0}) {new items};\band{draw,thick}}
  \caption{$M^\varepsilon$}\label{subfig:types-Meps}
\end{subfigure}
\caption{The segmentation of type space induced by the mechanisms $M$ and $M^\varepsilon$.}
\label{fig:segmentation}
\end{figure}

    Consider the segmentation of types associated with the mixed bundling mechanism $M$ depicted in \Cref{subfig:types-M}; in particular, consider the edge separating the region of types that choose the outside option $(0,0)$ from the region of types that choose the bundle $(1,1)$. We describe the construction in the coordinates $(s(\theta),\ell(\theta))$, where $s(\theta)=\theta_1-\theta_2$ is the position along the edge and $\ell(\theta)=\theta_1+\theta_2-p$ is the position in the orthogonal direction, i.e., type $\theta$'s utility from the bundle.

    The mechanism $M^\varepsilon$ is obtained from $M$ by adding to $\menu(M)$ new allocation-price pairs that form a  strictly convex curve in allocation-price space near the point $(\tfrac12,\tfrac12;\tfrac p2)$; see \Cref{subfig:menu-Meps}.
    In type space, the items along this curve are then bought by a neighborhood of types around the original edge; see \Cref{subfig:types-Meps}. The curve is parametrized by $s$ and constructed such that a type $\theta$ in the neighborhood buys the item with parameter $s(\theta)$ at price $p_{s(\theta)}$, where $|\tfrac p2-p_{s}|$ is on the order of $\varepsilon$. Types buying the same item hence sit on a line $s=\text{const}$ orthogonal to the original edge.

    $M^\varepsilon$ has a revenue loss on the order of $\varepsilon^2$ under the uniform distribution relative to the mixed bundling mechanism $M$. The width of the neighborhood is on the order of $\varepsilon$, and its length is asymptotically constant. Hence its area is on the order of $\varepsilon$. Up to a set of types with mass on the order of $\varepsilon^2$, half of the types in the neighborhood pay $p_s$ instead of $p$, whereas the other half of types pay $ p_s$ instead of $0$. Since $p-2p_s$ is on the order of $\varepsilon$, the total loss in revenue relative to $M$ is on the order of $\varepsilon^2$.

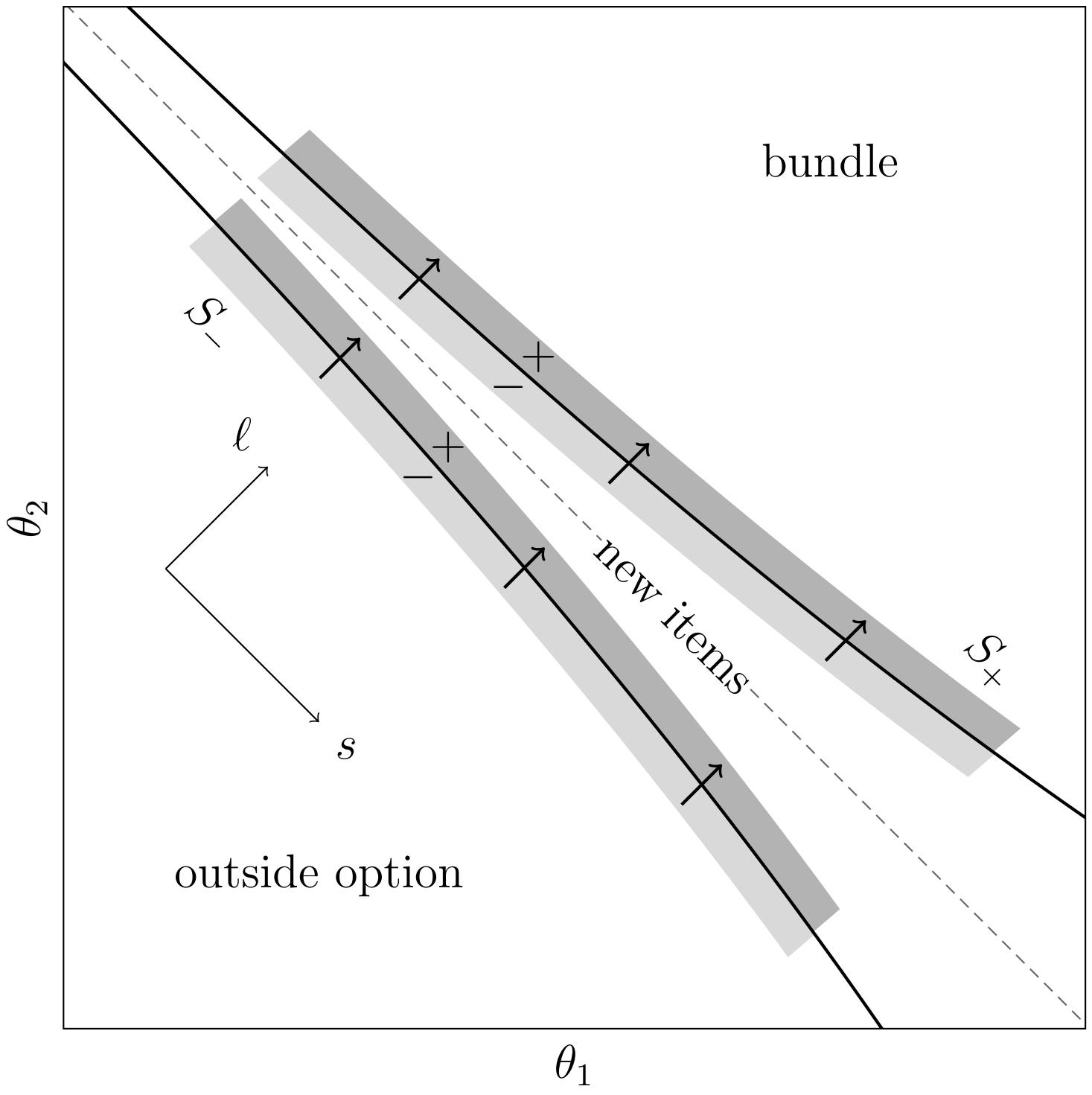
\begin{figure}[t]\centering\singlespacing
\tikzset{declare function={eps=1/90; p=(4-sqrt(2))/3; sleft=-0.350648; sright=0.442748; w=6*eps^(5/4);%
    phi(\s)=1+2.5*(\s+.8)^4; c(\s)=2*eps*phi(\s);
    tx(\s,\l)=(p+\l+sqrt(2)*\s)/2; ty(\s,\l)=(p+\l-sqrt(2)*\s)/2;},
  every plot/.style={variable=\s,samples=100}}
\newcommand\pt[2]{axis cs:{tx(#1,#2)},{ty(#1,#2)}}%
\newcommand\ribbon[5]{\path[#5] plot[domain=#1:#2] (\pt{\s}{#3}) -- plot[domain=#2:#1] (\pt{\s}{#4}) -- cycle;}
\newcommand\jump[2]{\draw[->,thick] (\pt{#1}{#2*c(#1)-.8*w})--(\pt{#1}{#2*c(#1)+.8*w});}
\begin{tikzpicture}\begin{axis}[scale only axis,width=.55\textwidth,height=.55\textwidth,axis lines=box,axis on top,
    xmin=.21,xmax=.65,ymin=.21,ymax=.65,xtick=\empty,ytick=\empty,xlabel={$\theta_1$},ylabel={$\theta_2$},label style={font=\normalsize}]
\draw[black!60,thin,densely dashed] (axis cs:0,{p})--(axis cs:{p},0);
\pgfplotsinvokeforeach{-1,1}{
  \ribbon{-.2}{.2}{#1*c(\s)-w}{#1*c(\s)}{fill=black!15}
  \ribbon{-.2}{.2}{#1*c(\s)}{#1*c(\s)+w}{fill=black!30}
  \node[inner sep=0pt] at (\pt{-.06}{#1*c(-.06)-.6*w}) {$-$};
  \node[inner sep=0pt] at (\pt{-.06}{#1*c(-.06)+.6*w}) {$+$};}
\ribbon{sleft}{sright}{-c(\s)}{c(\s)}{draw,thick}
\pgfplotsinvokeforeach{-.12,0,.12}{\jump{#1}{-1}\jump{#1}{1}}
\node[rotate=-45,anchor=north,inner sep=1pt] at (\pt{-.17}{-c(-.17)-1.3*w}) {$S_-$};
\node[rotate=-45,anchor=south,inner sep=1pt] at (\pt{.17}{c(.17)+1.3*w}) {$S_+$};
\node at (axis description cs:.25,.15) {outside option}; \node at (axis description cs:.75,.85) {bundle};
\node[fill=white,inner sep=.5pt,rotate=-45] at (\pt{.06}{0}) {new items};
\draw[->,thin] (axis description cs:.1,.45)--(axis description cs:.25,.3) node[below right] {$s$};
\draw[->,thin] (axis description cs:.1,.45)--(axis description cs:.2,.55) node[above left] {$\ell$};
\end{axis}\end{tikzpicture}
\caption{The mass transport $g^\varepsilon$. The display zooms in on the relevant region in \Cref{subfig:types-Meps}.}
\label{fig:transport}
\end{figure}

     The signed density $g^\varepsilon$ is constructed as follows. The neighborhood has a lower and an upper boundary, parametrized in the proof by curves $\ell=-c(s)$ and $\ell=c(s)$, respectively. Take two thin strips $S_-$ and $S_+$ of width on the order of $\varepsilon^{5/4}$ around $-c$ and $c$, respectively; see \Cref{fig:transport}.\footnote{The exponent $5/4$ of the width and the exponent $1/2$ of the weight $\sqrt{\varepsilon}=\varepsilon^{1/2}$ on the signed density $g^\varepsilon$ can be modified arbitrarily as long as both are positive, the former is larger than 1, and the sum of both exponents is strictly smaller than 2.} The signed density $g^\varepsilon$ is supported on the union of the two strips and negative below the curves $-c$ and $+c$ and positive above them.
     In effect, this means that $g^\varepsilon$ transports mass along every line $s=\text{const}$ through the strips, taking mass from below each curve and adding it above (indicated by the arrows in \Cref{fig:transport}).
     We further construct $g^\varepsilon$ such that it moves equal amounts of mass on the order of the width $\varepsilon^{5/4}$ across $-c$ and across $c$.
     The mass transport is favorable to the revenue of $M^\varepsilon$ because the payment jumps from $0$ to $p_s$ across $-c$ and from $p_s$ to $p$ across $c$. Since the two jumps add to $p$, the revenue gain along every line $s=\text{const}$ is $p$ times the amount of mass moved across either curve. Thus, $\int t_{M^\varepsilon} g^\varepsilon$ is on the order of $\varepsilon^{5/4}$.

    Finally, let us explain why any mechanism $M'$ of any given menu size $k\in\mathbb{N}$ is outperformed by $M^\varepsilon$ under $g^\varepsilon$ for sufficiently small $\varepsilon>0$. Since $M'$ has finite menu size, any two regions in type space where different options are bought are separated by a straight boundary line.
    The curve $\ell=c(s)$ is strictly convex, so any boundary line meets the strips $S_-$ and $S_+$ only in segments of vanishing length;
    the proof verifies that the set of positions $s$ at which some boundary line meets either strip has measure at most on the order of $\varepsilon^{1/8}$.
    By Fubini's theorem, think of $\int t_{M'} g^\varepsilon$ as a double integral over $s$ and $\ell$, with the inner integral taken over $\ell$ along the line $s=\text{const}$ orthogonal to the original edge. The inner integral at $s$ cancels whenever $t_{M'}$ is constant in $\ell$ across $S_-$ and across $S_+$, that is, whenever no boundary line of the cells of $M'$ meets either strip $S_\pm$ at $s$. Otherwise, the inner integral is at most on the order of the strip width $\varepsilon^{5/4}$. Hence, $\int t_{M'} g^\varepsilon$ is at most on the order of $\varepsilon^{11/8}=\varepsilon^{5/4}\cdot\varepsilon^{1/8}$, which vanishes faster than $\int t_{M^\varepsilon} g^\varepsilon\approx\varepsilon^{5/4}$.
    The constants in these revenue bounds depend on $M'$ only through its menu size $k$, so $M^\varepsilon$ outperforms all mechanisms of menu size $k$ for sufficiently small $\varepsilon>0$.

    The extension to non-uniform distributions is straightforward. First, note that smooth, non-uniform densities that are bounded from above and below only introduce constants to the calculations and do not affect speeds of convergence. Second, such densities can approximate all other probability measures in the topologies we consider, hence suffice for the proof.
    Third, the perturbation arguments described above apply to any optimal mechanism $M$ with finite menu size, not just mixed bundling.

\begin{remark}[Independent Types]
    The perturbed density $f^\varepsilon=f+\sqrt{\varepsilon}g^\varepsilon$ constructed in the proof is not necessarily a product density if $f$ is a product density. Thus, the present construction does not directly imply a genericity result under the assumption of independent types. However, we have strong reasons to believe that our arguments can be modified to account for this additional structure on type distributions.
\end{remark}

\begin{remark}[Higher Dimensions]\label{higher-dimensions}
    The arguments can be extended to higher-dimensional screening problems by carrying out all perturbations in two dimensions only. To see this, recall that we have described the construction in the coordinates $s(\theta)$ and $\ell(\theta)$, that is, the position along an edge of indifferent types between two items and the utility difference between these items, respectively. In higher dimensions, the edge becomes a hyperplane, $s(\theta)$ becomes the position along a fixed direction tangential to the hyperplane, and $\ell(\theta)$ remains the position along the direction orthogonal to the hyperplane. The curve of new items in allocation-price space is then defined verbatim in terms of $s$ and $\ell$. The new items are again chosen by types in a neighborhood of the hyperplane with curved boundaries. The signed density can be defined analogously along strips around these curved boundaries. A formalization would yield no additional insights and is hence omitted.
\end{remark}

\begin{remark}[Stronger Topologies]
    The strongest topology for which we  establish denseness is the topology of uniform convergence on the space $\mathcal{F}$ of continuous densities. The present construction leaves open whether denseness can be achieved in stronger topologies. To see this, recall that the perturbation term $\sqrt{\varepsilon}g^\varepsilon$ constructed in the proof is supported on two strips of width on the order of $\varepsilon^{5/4}$ (see \Cref{fig:transport}). As $\varepsilon\to 0$, $\sqrt{\varepsilon}g^\varepsilon$ converges to zero uniformly but the maximum norm of its  gradient is on the order of $\sqrt{\varepsilon}/\varepsilon^{5/4}=\varepsilon^{-3/4}$ and hence diverges.
    Thus, the perturbation term does not vanish in the $C^1$-norm. We leave the question of whether our arguments can be extended to stronger topologies for future research.
\end{remark}

\section{Costs and Other Frictions}
\label{sec:discussion}

We have established the genericity of complex menus under the implicit assumption that adding options to a menu is free. In this section, we ask: How do frictions to product differentiation affect how finely a seller differentiates? What happens when product differentiation is facilitated by technological progress, such as advances in information technology?

We explore four reasons why adding new options to a menu might be costly or difficult. First, each additional option on the menu may involve a setup cost, e.g., of designing and creating the production capabilities for an additional version of a product, administering it, and marketing it to consumers (\Cref{subsec:menucosts}). Second, large menus may be costly for buyers to navigate, translating into shadow costs for the seller (\Cref{subsec:cognitive}). Third, there may be technological constraints that render some allocations infeasible (\Cref{subsec:discrete}). Fourth, there may be variable costs; in particular, costs may scale favorably with the size or quality of allocations (\Cref{subsec:varicosts}). These frictions, if sufficiently strong, can all rationalize simple menus. The question is hence whether the conclusion of \Cref{thm:master} is an artifact of the absence of frictions.

For generic type distributions, we show that the optimal menu size grows without bound as these frictions vanish. Technological progress that reduces these frictions may hence increase the extent to which sellers price discriminate through vertical differentiation. As discussed in the introduction, this effect may be particularly pronounced in the context of digital goods.

\subsection{Menu-Size Costs}

\label{subsec:menucosts}
Suppose a menu of size $k$ costs the seller $\lambda C(k)$, where $\lambda\geq 0$ is a scaling parameter and $C$ is an increasing function of menu size with $C(1)=0$. The seller's profit is given by
$
    \Pi_\lambda(M;\mu)=\REV(M;\mu)-\lambda C(\lvert\menu(M)\rvert).
$
To ensure the existence of a profit-maximizing mechanism, we assume for simplicity that $C$ is unbounded and that infinite menus are not feasible unless $\lambda=0$.
We understand menu-size costs broadly, as covering product design, creation of production capabilities, administration of the menu and optimal pricing, as well as product-specific regulatory filings.

We refer to the smallest menu size among profit-maximizing mechanisms
$$
k_\lambda(\mu):=\min \left\{\lvert\menu(M)\rvert:\: M\in \argmax_{\tilde M\in\mathcal{M}} \Pi_\lambda(\tilde M;\mu)\right\}
$$
as the \textbf{optimal menu size}. We shall use the same term with the obvious modifications in the subsequent subsections as well.

The following lemma shows that, as frictions due to menu-size costs vanish, the optimal menu size monotonically approaches that of the frictionless benchmark.

\begin{lemma}
\label{lemma:menu_costs}
    For every type distribution $\mu\in\mathcal{P}$, the optimal menu size $k_\lambda(\mu)$ is non-increasing in $\lambda$ and $k_\lambda(\mu)\to k_0(\mu)$ as $\lambda\to 0$.
\end{lemma}

\begin{proof}
    See \Cref{subsec:frictions}.
\end{proof}

We immediately conclude from \Cref{thm:master} and \Cref{lemma:menu_costs} that small menu-size costs generically lead to large optimal menus.

\begin{corollary}
\label{cor:menu_costs}
    For generic type distributions, the optimal menu size increases without bound as $\lambda\to 0$.
\end{corollary}

As we have discussed in the introduction, menu-size costs are likely to fall due to technological progress, in particular, progress in information technology. But technological progress is not the only driver of decreasing menu-size costs. As we measure these costs per transaction, an increase in effective market size, e.g.,  due to globalization, has a similar effect, as the costs are simply spread across more transactions.

We remark that while lower menu-size costs never hurt the seller, they need not benefit buyers. An additional option can allow the seller to serve buyers that the seller would otherwise exclude. However, it can also allow the seller to raise the price of its premium product. As a consequence, both consumer surplus and total surplus (even gross of menu-size costs) may fall as optimal menus grow in size.

\begin{example}
\label{example:setup_costs}
Suppose a share $\tfrac{7}{10}$ of buyers have values $(\tfrac{4}{5},\tfrac{1}{5})$, that the remaining share $\tfrac{3}{10}$ of buyers have values $(1,1)$, and that $C(k)=k-1$. For scaling parameter $\lambda\in (\tfrac{1}{10},1)$, the optimal menu offers the premium version $(1,1)$ at price $1$, and all buyers buy it. For $\lambda\in (0,\tfrac{1}{10})$, i.e., as menu-size costs fall, the seller additionally offers the basic version $(1,0)$ at price $\tfrac{4}{5}$ to the first group of buyers and raises the price of the premium version to $\tfrac{9}{5}$. Consumer surplus  falls from $\tfrac{3}{10}$ to $\tfrac{3}{50}$, and total surplus gross of menu-size costs falls from $\tfrac{13}{10}$ to $\tfrac{29}{25}$. The calculations are in \Cref{app:examples}.
\end{example}

\subsection{Buyers' Cognitive Costs}
\label{subsec:cognitive}
\label{subsec:buyer-cognitive}

Large menus can also be costly for buyers. We consider a simple model where inspecting a menu of size $k$ comes with a cognitive cost $\lambda C(k)$ for buyers, with $\lambda$ and $C$ as in \Cref{subsec:menucosts}. Cognitive costs may involve the effort of examining the menu, discovering one's valuations, and identifying the utility-maximizing option.
Before inspecting a menu, each buyer holds a correct belief $\beta\in\mathcal{P}$ about their type.
The population distribution $\tau$ of beliefs is known to the seller and averages to the type distribution $\mu$, that is, $\int\beta\,d\tau(\beta)=\mu$. A buyer inspects a menu if and only if their expected surplus covers the cognitive cost. If the buyer inspects, they learn their type and then choose their favorite option from the menu.

We deliberately keep the model simple by restricting the buyer's inspection choice to a single binary decision. We thereby isolate the extensive margin of cognitive costs, that a buyer may be deterred from buying at all when confronted with a large menu. This choice overload effect has been documented in the context of consumer choice (e.g., \citet{iyengar2000choice}; \citet{chernev2015choice}).

The following lemma shows that if a given menu size is too small to be optimal without cognitive costs, then it is also too small to be optimal for all belief distributions $\tau$ once cognitive costs are small enough. Analogously to \Cref{subsec:menucosts}, for $\tau$ averaging to $\mu$, we denote by  $k_\lambda(\tau)$ the optimal menu size at cost scale $\lambda>0$ and by $k_0(\mu)$ the optimal menu size in the absence of cognitive costs.

\begin{lemma}
\label{lemma:attention}
Let $\mu\in\mathcal{P}$. For every $\lambda>0$ and every belief distribution $\tau$ averaging to $\mu$, an optimal mechanism exists. Moreover, for every integer $k\leq k_0(\mu)$ there exists $\bar\lambda>0$ such that, for all $\lambda\leq\bar\lambda$ and all belief distributions $\tau$ averaging to $\mu$, the optimal menu size is at least $k$.
\end{lemma}

\begin{proof}
See \Cref{subsec:frictions}.
\end{proof}

As with menu-size costs, \Cref{thm:master} and \Cref{lemma:attention} imply that optimal menus are generically large when cognitive costs are small.

\begin{corollary}
\label{cor:attention}
For
generic type distributions $\mu$ and every belief distribution $\tau$ averaging to $\mu$, the optimal menu size $k_\lambda(\tau)$ under cognitive costs tends to infinity as $\lambda\to 0$.
\end{corollary}

Unlike a higher menu-size cost, a higher cognitive cost can increase the optimal menu size. The idea is that a seller must leave every buyer who inspects the menu enough expected surplus to cover the increasing cognitive cost. The seller can do so (i) by lowering the menu size or (ii) by serving only buyers with high expected valuations. In the latter case, it can then be optimal for the seller to discriminate more finely among the selected subset of high-value buyers.

\begin{example}
\label{example:cognitive_costs}
Suppose there are three groups of buyers with shares $(\tfrac{1}{10},\tfrac{2}{10},\tfrac{7}{10})$ and values $(1,0)$, $(0,\tfrac{1}{2})$, and $(\tfrac{1}{8},\tfrac{1}{8})$, respectively, and $C(k)=k-1$. For simplicity, suppose buyers perfectly know their preferences. For cost parameter $\lambda\in(0,\tfrac{1}{8})$, the optimal menu offers the bundle $(1,1)$ at price $\tfrac{1}{4}-\lambda$, and all buyers buy it. For $\lambda\in(\tfrac{1}{8},\tfrac{1}{6})$, it is optimal to offer instead $(1,0)$ at price $1-2\lambda$ and $(0,1)$ at price $\tfrac{1}{2}-2\lambda$; the first group buys $(1,0)$, the second group buys $(0,1)$, and the third group  takes the outside option. Thus, as cognitive costs increase, the optimal menu size increases from two to three (when counting the outside option). The larger menu does better at higher cognitive costs because the cost is now incurred only by buyers with high values. The calculations are in \Cref{app:examples}.
\end{example}

\subsection{Technological Constraints}
\label{subsec:discrete}

Suppose the seller is restricted to allocations in a closed set $Q \subseteq
[0,1]^2$, with $(0,0)\in Q$. We interpret allocations outside of $Q$ as technologically infeasible.
Two features of the allocation space $Q$ determine whether large menus are generically optimal: whether the seller can differentiate along more than one dimension, and how finely.

First, suppose the seller can differentiate products along a single allocation dimension only, $Q = \{(r,r) : r \in [0,1]\}$. In this case, the seller's problem collapses to a one-dimensional screening problem. By the classical results of \citet{myerson1981optimal} and \citet{riley1983optimal} (see also \citet{stokey1979intertemporal}), it is optimal for the seller to offer only the best option $(1,1)$ with no further quality differentiation. Hence, a minimal degree of variety in allocations is necessary for revenue-maximizing menus to be complex in our model.

By contrast, if the allocation space $Q$ is convex with non-empty interior, so that the seller can differentiate along two dimensions, then the proof of \Cref{thm:master} goes through verbatim. For instance, taking $Q$ to be the two-dimensional probability simplex yields the genericity of infinite menus for the unit-demand monopoly problem studied in \citet{thanassoulis2004haggling}
 and subsequent papers.

Second, suppose fine product differentiation is impossible, i.e., $|Q|$ is small. Large menus are then trivially infeasible. Moreover, it is not true that generic type distributions require optimal menu size exactly $|Q|$. This is because the available allocations may not allow the seller to separate certain sets of types for which separation would be profitable; compare with the proof sketch in \Cref{sec:result}.

By contrast, as the allocation space becomes richer, the seller regains the ability to separate types as needed, leading to more complex menus. For example, the seller may find ways to degrade product features or recombine existing product features more easily.
Let
$$
  \mesh(Q) = \sup_{q \in [0,1]^2} \ \inf_{q' \in Q} \ \|q - q'\|_1
$$
measure the coarseness of the allocation space, e.g., $\mesh(\{0, \tfrac{1}{n},
\ldots, 1\}^2) = \tfrac{1}{n}$.

\begin{corollary}
\label{cor:discrete}
For
generic type distributions%
, the optimal menu size along any sequence of allocation spaces $(Q_n)_{n\in\mathbb{N}}$ tends to infinity as $\mesh(Q_n) \to 0$.
\end{corollary}

\begin{proof}
    See \Cref{subsec:frictions}.
\end{proof}

\subsection{Production Costs}
\label{subsec:varicosts}

Suppose the seller incurs a cost $C(q)$ for every buyer who obtains allocation $q\in Q=[0,1]^2$. We assume throughout that cost functions $C: Q\to\mathbb{R}_+$ are Lipschitz continuous, satisfy $C(0)=0$, and allow gains from trade, i.e., $C(q)<q_1+q_2$ for some $q\in Q$.
A mechanism $M$ is now \textbf{optimal} if it maximizes the seller's profit
$$
\Pi_C(M;\mu)=\int_\Theta\bigl[t_M(\theta)-C(x_M(\theta))\bigr]\,d\mu(\theta).%
$$

A key determinant of whether or not the seller finds it optimal to offer large menus is how cost-effective it is to offer inferior products.
Formally, we say that the cost function $C$ admits \textbf{cost-effective downgrades} if for each allocation $q\in Q\setminus\{\mathbf{0}\}$ there exists $r\in(0,1)$ such that $C(rq)\leq rC(q)$. If the last inequality in this definition is strict ($C(rq)< rC(q)$), we say that $C$ admits
\textbf{strictly cost-effective downgrades}. In plain words, a downgrade is cost-effective if the surplus at $rq$ is no more than a factor of $r$ lower than at $q$. The standard specification of (strictly) convex costs (e.g., \citet{mussa1978monopoly}; \citet{rochet1998ironing}) admits (strictly) cost-effective downgrades. Moreover, the same is true for any homogeneous-of-degree-1 cost function, e.g., $C(q)=\min\{q_1,q_2\}$, or, more generally, any cost function that exhibits non-decreasing ray-average costs in the sense of \citet{baumol1977proper}, i.e., $r\mapsto C(rq)/r$ is non-decreasing for all $q\in Q$.

\begin{corollary}
    \label{cor:downgrades}
    \Cref{thm:master} extends to cost functions that admit cost-effective downgrades. If the cost function admits strictly cost-effective downgrades and the type distribution admits a continuous density bounded away from zero, then every optimal mechanism has infinite menu size.
\end{corollary}

The proof is a straightforward extension of the arguments behind \Cref{thm:master}. There, we have perturbed a given finite menu by adding a continuum of new options in a neighborhood of a scaled-down option $(rq;rp)$ of some option $(q;p)$ that is adjacent to the outside option (see \Cref{fig:menus,fig:segmentation}); the proof takes $r=\tfrac12$, but any scalar $r\in(0,1)$ can be used instead. The proof of \Cref{thm:master} shows that the revenue loss from this perturbation is of second order. If $C$ admits cost-effective downgrades, then for $r$ such that $C(rq)\leq rC(q)$ the increase in costs is at most of second order by the same arguments (assuming, as above, that $C$ is Lipschitz continuous). Indeed, if $C$ admits strictly cost-effective downgrades, then there are first-order cost savings, so the perturbation increases profit. If there are no first-order cost savings, then the proof of \Cref{cor:downgrades} is completed by the same steps as in the proof of \Cref{thm:master}; we omit the formal arguments.

If the cost function does not admit cost-effective downgrades, then it may be relatively more costly for the seller to offer inferior products.
There are hence two counteracting effects: the seller's motive to screen buyer types by offering differentiated products, as established in \Cref{thm:master}, and the seller's motive to save costs.
Which effect is stronger depends on the type distribution as well as the size and curvature of costs.

However, if the seller can randomize allocations, for instance by offering rationed products at a discount, then the seller effectively convexifies their cost function $C$, and so we immediately conclude from \Cref{cor:downgrades}:

\begin{corollary}
\label{cor:random}
 If the seller can offer randomized allocations, then \Cref{thm:master} extends to all cost functions.
\end{corollary}

\section{Conclusion}

We prove that optimal screening menus are infinitely large in generic multidimensional screening problems, even in settings in which efficient contracts are very simple. Screening considerations alone are sufficient for this complexity to arise; other factors such as production costs or preferences are not needed as an explanation. The popular single-dimensional model---which abstracts away from the multidimensionality inherent in most real-world situations---fails to capture this source of complexity. Indeed, as \citet{stokey1979intertemporal}, \citet{myerson1981optimal}, and \citet{riley1983optimal} show, in linear single-dimensional settings optimal screening can be achieved with just a single offer.

We posit that complexity costs are the reason why we frequently do not observe this complexity in practice.
The generic need for large menus we establish arises in canonical multi-dimensional screening models, which abstract from complexity costs and other frictions. Our analysis therefore suggests that complexity costs are essential for explaining simple menus. These costs might arise on both the seller and the buyer side. On the seller side, they reflect the costs of designing and creating the production capabilities for new varieties, as well as of administering, pricing, and marketing complex menus. On the buyer side, complexity costs reflect the cognitive costs of choosing from a large menu.

These points are timely because recent and rapid technological progress has lowered complexity costs in some industries, and this may be particularly relevant in the context of digital goods. In cloud computing, for instance, we already see highly complex screening menus. A further decrease in complexity costs on both the seller and buyers' side, a likely consequence of continued technological progress, might hence lead to a world in which buyers will be screened in a highly complex way in an increasing number of contexts; indeed, they might be expected to and forced to rely on AI agents in their choice processes.

The welfare consequences of these developments will depend on the specific industry. As we show, sellers unambiguously benefit from lower complexity costs. Buyers, however, may gain or lose from these developments, and total surplus may even fall, depending on further particulars of the situation.
An optimal regulatory response hence also depends on these further particulars.
In particular, regulators or platforms might
optimally choose to restrict a seller's ability to offer complex menus, e.g., by charging a fee per offered item, even if there are no underlying complexity costs.

While we focus on monopolistic screening, our results suggest
that similar complexity might arise in oligopolistic settings with sufficient horizontal differentiation among sellers. Our results can be applied if the revenue that an oligopolist earns from its captive consumers is sufficiently large relative to the revenue from the part of the market where the oligopolist competes with others.\footnote{This follows from the continuity results in \Cref{subsec:openness}.
In the absence of any horizontal differentiation between sellers, the Bertrand logic would prevail and the screening-driven complexity disappears.}   In this case, the generic benefits of complex menus in monopolistic screening outweigh the potential losses due to competitive forces.
The menu-size complexity we focus on is distinct from and complementary to the obfuscation strategies studied by the rich literature on market competition, see, e.g., \citet{Heidhues-Koszegi-QJE-2017}, \citet{Ellison-Ellison-ECTA-2018}, and \citet{Fehr-Wu-2025}.

	\appendix
\crefalias{section}{appendix}
\crefalias{subsection}{appendix}

	\section{Proofs}
	\label{sec:proof}

We write
	\begin{align*}
    \REV^*(\mu)&=\sup_{M\in\mathcal{M}}\REV(M;\mu),\\
		\mathcal{M}^*(\mu)&=\{M\in\mathcal{M}:\REV(M;\mu)=\REV^*(\mu)\},\\
		\mathcal{M}^{\leq k}&=\{M\in\mathcal{M}: \lvert\menu(M)\rvert\leq k\},\\
		\REV^{\leq k}(\mu)&=\sup_{M\in\mathcal{M}^{\leq k}}\REV(M;\mu),\\
		\mathcal{P}^{>k}&=\{\mu\in\mathcal{P}:\mathcal{M}^*(\mu)\cap\mathcal{M}^{\leq k}=\emptyset\}.
	\end{align*}
    In words, $\REV^*(\mu)$ is the optimal revenue, $\mathcal{M}^*(\mu)$ is the set of optimal mechanisms, $\mathcal{M}^{\leq k}$ is the set of mechanisms of menu size at most $k$, $\REV^{\leq k}(\mu)$ is the optimal revenue among mechanisms of menu size at most $k$, and $\mathcal{P}^{>k}$ is the set of type distributions for which all optimal mechanisms have menu size greater than $k$.
	If $\mu$ has density $f$, we analogously write $\REV^*(f)$, $\REV^{\leq k}(f)$, and $\mathcal{M}^*(f)$, and we use the same notation when $f$ is replaced by an integrable signed density. The sets $\bar{\mathcal{F}}^{>k}$ and $\mathcal{F}^{>k}$ are defined analogously to $\mathcal{P}^{>k}$ with $f$ in place of $\mu$ and $\bar{\mathcal{F}}$ and $\mathcal{F}$ in place of $\mathcal{P}$, respectively.

A $G_\delta$ set is a countable intersection of open sets, so a set is generic if it contains a dense $G_\delta$ set.

\subsection{Openness}
	\label{subsec:openness}

	\begin{lemma}
		\label{lemma:rev_continutity}
		For each $k\in\mathbb{N}$, $\mu\mapsto\REV^*(\mu)$ and $\mu\mapsto\REV^{\leq k}(\mu)$ are continuous on $\mathcal{P}$.
	\end{lemma}

	\begin{proof}
	By \citet[Theorem~6(ii)]{hart2025root}, optimal revenue over any class of mechanisms that is closed under rescaling of menu prices is continuous with respect to the Wasserstein distance. The Wasserstein distance here metrizes the weak topology since the type space $\Theta$ is compact. The class of mechanisms with menu size at most $k$ is closed under rescaling of menu prices.
	\end{proof}

    \begin{lemma}
	\label{lemma:bounded-attainment}
	For every $k\in\mathbb{N}$ and every $\mu\in\mathcal{P}$, the optimal revenue $\REV^{\leq k}(\mu)$ among mechanisms offering at most $k$ items is attained.
    \end{lemma}

\begin{proof}[Proof of \Cref{lemma:bounded-attainment}]
    Let $(M^n)_{n\in\mathbb{N}}\subset\mathcal{M}$ be a sequence of IC and IR mechanisms with $\lvert\menu(M^n)\rvert\leq k$ for all $n\in\mathbb{N}$ and such that $\REV(M^n;\mu)\to \REV^{\leq k}(\mu)$. For each $n\in\mathbb{N}$, embed $\menu(M^n)$  in $([0,1]^2\times [0,2])^k$ by arbitrary enumeration, repeating an item if necessary. (Prices are in $[0,2]$ by IR.) Since $([0,1]^2\times [0,2])^k$ is compact, $\menu(M^n)$ converges up to a subsequence; let $M\in\mathcal{M}$ be defined by the buyer's rational choice from $\lim_{n\to\infty} \menu(M^n)$, breaking ties in favor of the highest price. By Proposition 5 in \citet{hart2026existence}, $\lim_{n\to\infty} \REV(M^n;\mu)\leq \REV(M;\mu)$, completing the proof.
\end{proof}

\begin{proof}[Proof of \Cref{thm:master}: Openness]
    By \Cref{lemma:bounded-attainment}, every optimal mechanism for $\mu\in\mathcal{P}$ has menu size greater than $k$ if and only if $\REV^*(\mu)>\REV^{\leq k}(\mu)$. Thus,
	$
	\mathcal P^{>k}=\{\mu\in\mathcal P:\REV^*(\mu)>\REV^{\leq k}(\mu)\}.
	$
    By \Cref{lemma:rev_continutity}, $\mathcal{P}^{>k}$ is open. Both $\bar{\mathcal{F}}$ and $\mathcal{F}$ are subsets of $\mathcal{P}$ when identifying densities with their probability measures. Hence, $\bar{\mathcal{F}}^{>k}$ and $\mathcal{F}^{>k}$ are open in $\bar{\mathcal{F}}$ and $\mathcal{F}$, respectively, because these spaces are equipped with stronger topologies than the weak topology.
\end{proof}

\begin{remark}[Infinite Menu Size and Openness]
        \label{remark:notopen}
        The set of type distributions for which every optimal mechanism has infinite menu size is not open in $\mathcal{P}$. To see this, note that the finitely supported type distributions are dense in $\mathcal{P}$ and that, for a type distribution with finite support, there exists an optimal mechanism with finite menu size. We suspect that the same is true in $\mathcal{F}$ and $\bar{\mathcal{F}}$ but have not constructed explicit examples.
\end{remark}

\subsection{Denseness}
\label{subsec:denseness}

\begin{lemma}
\label{lemma:cap}
Fix $k \ge 2$ and let $f \in C^\infty(\Theta)$ be a strictly positive probability
density such that $\REV^{\le k}(f) = \REV^*(f)$. Then there exist constants
$A > 0$, $D, C_k \ge 0$, and $\varepsilon_0 > 0$ such that, for every
$\varepsilon \in (0, \varepsilon_0)$ and $\gamma = \varepsilon^{5/4}$, there
exist $M^{\varepsilon} \in \mathcal{M}$ and
$g^{\varepsilon} \in C^{\infty}(\Theta)$ satisfying
\begin{equation}
 \int g^\varepsilon = 0, \quad \|g^\varepsilon\|_\infty \leq 1,
 \label{eq:cap-mass}
\end{equation}
\begin{equation}
 \REV(M^\varepsilon; f) \geq \REV^*(f) - D\varepsilon^2,
 \label{eq:cap-loss}
\end{equation}
\begin{equation}
 \REV(M^\varepsilon; g^\varepsilon) = A\gamma, \qquad
 \REV^{\leq k}(g^\varepsilon) \leq C_k \gamma^{3/2} \varepsilon^{-1/2}.
 \label{eq:cap-competitor}
\end{equation}
In particular, there exists $\bar\varepsilon \in (0, \varepsilon_0]$ such that,
for every $\varepsilon \in (0, \bar\varepsilon)$, the function
$f^{\varepsilon} = f + \sqrt{\varepsilon}\, g^{\varepsilon}$ is a probability
density in $C^{\infty}(\Theta)$ with
$\|f^{\varepsilon} - f\|_{\infty} \le \sqrt{\varepsilon}$ and
\begin{equation}
 \REV^*(f^{\varepsilon}) \;>\; \REV^{\le k}(f^{\varepsilon}).
 \label{eq:cap-conclusion}
\end{equation}
\end{lemma}

\begin{proof}

Select an arbitrary mechanism $M \in \mathcal{M}^*(f) \cap \mathcal{M}^{\le k}$, which exists by \Cref{lemma:bounded-attainment}.
For an item $(q;p) \in \menu(M)$, define $u(q;p)(\theta) = q\cdot\theta - p$ and the polyhedral cell
$$
C(q;p) = \{\theta \in \Theta : u(q;p)(\theta) \ge u(q';p')(\theta) \text{ for all } (q';p') \in \menu(M)\}.
$$
Two items of $\menu(M)$ are \textbf{adjacent} if their cells have non-empty interior and the intersection of their cells is a
non-degenerate line segment.

\paragraph{Step 1: selecting an edge.}

 Optimality requires that $M$ allocates the
outside option $(\mathbf{0};0)$ to a set of types $C(\mathbf{0};0)$ of positive
measure (e.g., \citet[Theorem~16]{manelli2007multi-dimensional}).
Select
$(q;p) \in \menu(M)$ adjacent to $(\mathbf{0};0)$ and
$\theta^0 \in \relint\bigl(C(\mathbf{0};0) \cap C(q;p)\bigr)$.\footnote{Any pair of adjacent items with distinct prices would suffice for the proof.} Optimality
requires $p > 0$.
Let $e$ be a unit vector with $e\cdot q = 0$ and $e_i>0$ if $q_i=0$, and define the invertible affine
map $T(\theta) = (s(\theta), \ell(\theta))$ by
$$
s(\theta) := e\cdot(\theta - \theta^0), \qquad
\ell(\theta) := u(q;p)(\theta) = q\cdot\theta - p.
$$
In words, $s(\theta)$ is the position along the edge $C(\mathbf{0};0)\cap C(q;p)$, measured relative to $\theta^0$, and $\ell(\theta)$ is the position in the direction orthogonal to the edge, measured in utils.
In particular, $\ell(\theta)$ vanishes exactly on the line through the edge
$C(\mathbf{0};0) \cap C(q;p)$, and type $\theta$ prefers $(q;p)$ to
$(\mathbf{0};0)$ if and only if $\ell(\theta) \ge 0$.
Choose $\bar s, \bar \ell > 0$ such that
$$
R = \{\theta\in\Theta : |s(\theta)| \le \bar s,\ |\ell(\theta)| \le \bar\ell\} \subset
\interior\bigl(C(\mathbf{0};0) \cup C(q;p)\bigr).
$$

\paragraph{Step 2: constructing $(M^\varepsilon)_\varepsilon$.}
Let $\phi:s(\Theta)\to\mathbb{R}$ be any smooth, strictly convex function with $1/B<\phi,\phi',\phi''<B$ for some $B>0$.

For $\sigma\in s(\Theta)$, define the menu item $(q_\sigma;p_\sigma)$ with
$$
q_\sigma=\tfrac12 q+\varepsilon\phi'(\sigma)e \qquad \text{and} \qquad  p_\sigma=\tfrac12 p-\varepsilon\phi(\sigma)+\varepsilon\phi'(\sigma)(e\cdot\theta^0+\sigma),
$$ and
let $M^\varepsilon$ be given by rational choice from
$\menu(M)\cup\{(q_\sigma;p_\sigma):\sigma\in s(\Theta)\}$. $M^\varepsilon$ is well-defined
because $s(\Theta)$ is compact and $(q_\sigma;p_\sigma)$ is continuous in $\sigma$. Moreover, for all sufficiently small $\varepsilon>0$, $p_\sigma\in (0,p)$ and
$q_\sigma\in[0,1]^2$ since $e_i>0$ if $q_i=0$.
For $\sigma\in s(\Theta)$, we have
$$
u(q_\sigma;p_\sigma)(\theta)=\tfrac12\ell(\theta)+\varepsilon\bigl(\phi(\sigma)+\phi'(\sigma)(s(\theta)-\sigma)\bigr).
$$

Out of the new items $(q_\sigma;p_\sigma)_{\sigma\in s(\Theta)}$, type
$\theta\in\Theta$ prefers the option indexed by $s(\theta)$.
This is because $\phi$ is strictly convex, so comparing the bracketed terms in the previous display gives
$\phi(s(\theta))-\phi(\sigma)-\phi'(\sigma)(s(\theta)-\sigma)\ge0$
with equality if and only if $\sigma=s(\theta)$. The maximized utility from the new items is
therefore
$$
\max_{\sigma\in s(\Theta)}u(q_\sigma;p_\sigma)(\theta)=\tfrac12\ell(\theta)+\varepsilon\phi(s(\theta)).
$$

Type $\theta$ prefers the new items to
$\{(\mathbf{0};0),(q;p)\}$ if and only if
$\tfrac12\ell(\theta)+\varepsilon\phi(s(\theta))\geq\max\{0,u(q;p)(\theta)\}=\tfrac12(\ell(\theta)+|\ell(\theta)|)$; that is, if and only if $|\ell(\theta)|\leq c(s(\theta))$ with $c(s)=2\varepsilon\phi(s)$.

Note that, for almost every $\theta\in R\subset \interior(C(\mathbf{0};0)\cup C(q;p))$, the payment is given by
$$
t_{M^\varepsilon}(\theta)=
\begin{cases}
0&\text{if }\ell(\theta)<-c(s(\theta)),\\
p_{s(\theta)}&\text{if }|\ell(\theta)|<c(s(\theta)),\\
p&\text{if }\ell(\theta)>c(s(\theta)).
\end{cases}
$$

\paragraph{Step 3: bounding the revenue loss of $M^\varepsilon$ under $f$.}
Let $u(\theta)$ and $u^\varepsilon(\theta)$ denote the indirect utility functions of
$M$ and $M^\varepsilon$. Let
$\psi(\theta)=u^\varepsilon(\theta)-u(\theta)$. Since
$M^\varepsilon$ is given by rational choice from a superset of $\menu(M)$, $\psi\ge0$.

We first bound $\psi$. For every $\theta\in\Theta$,
$u(\theta)\ge\max\{0,u(q;p)(\theta)\}=\tfrac12(\ell(\theta)+|\ell(\theta)|)$,
and $u^\varepsilon(\theta)>u(\theta)$ only if a new item is chosen at
$\theta$, so
$$
\psi(\theta)
\le\max\Bigl\{0,\max_{\sigma\in s(\Theta)}u(q_\sigma;p_\sigma)(\theta)-\tfrac12\bigl(\ell(\theta)+|\ell(\theta)|\bigr)\Bigr\}=\max\Bigl\{0,\varepsilon\phi(s(\theta))-\tfrac12|\ell(\theta)|\Bigr\}.
$$
Because $\phi\leq B$, we have $\|\psi\|_\infty\le B\varepsilon$, and the set
$\{\psi>0\}\subset\{\theta\in\Theta:|\ell(\theta)|<2B\varepsilon\}$ has Lebesgue measure on the order of $\varepsilon$.

By Proposition 2 in \citet{rochet1987necessary},
$t_{M^\varepsilon}(\theta)=\theta\cdot\nabla u^\varepsilon(\theta)-u^\varepsilon(\theta)$
and $t_M(\theta)=\theta\cdot\nabla u(\theta)-u(\theta)$ for almost every
$\theta\in\Theta$. Thus,
$$
\REV(M^\varepsilon;f)-\REV(M;f)=\int_\Theta\bigl(\theta\cdot\nabla\psi(\theta)-\psi(\theta)\bigr)f(\theta)\,d\theta.
$$

Applying the divergence theorem, e.g., as in \citet[Lemma~2]{rochet1998ironing}, gives
$$
\REV(M^\varepsilon;f)-\REV(M;f)=\int_{\partial\Theta}\psi(\theta)f(\theta)(\theta\cdot n(\theta))\,d\theta
-\int_\Theta\psi(\theta)\bigl(3f(\theta)+\theta\cdot\nabla f(\theta)\bigr)\,d\theta,
$$
where $n(\theta)$ is the outward unit normal to the boundary $\partial\Theta$ at $\theta$.

Inequality~\eqref{eq:cap-loss} can be seen from the previous display.
The surface integral is non-negative because $\psi f\ge0$ and
$\theta\cdot n(\theta)\geq 0$.
The volume integral is on the order of $-\varepsilon^2$ because $\|\psi\|_\infty\le B\varepsilon$,
$\{\psi>0\}\subset\{\theta\in\Theta:|\ell(\theta)|<2B\varepsilon\}$ has Lebesgue measure on the order of $\varepsilon$, and the hazard term $3f(\theta)+\theta\cdot\nabla f(\theta)$ is uniformly bounded by the compactness of $\Theta$.

\paragraph{Step 4: constructing $(g^\varepsilon)_\varepsilon$.}

The following functions $\chi$ and $h$ are chosen independently of $\varepsilon$ and are used in the construction of $g^\varepsilon$. First,
choose an arbitrary function $\chi\in C^\infty(\mathbb{R})$ with $0\le\chi\le1$, $\chi=0$ outside
$[-\bar s,\bar s]$, and $\int\chi>0$. Second, choose an odd function $h\in C^\infty(\mathbb{R})$, i.e.,
 $h(-z)=-h(z)$ for all $z\in\mathbb{R}$, with $h=0$ outside $(-1,1)$,
$\|h\|_\infty\le1$, $h\ge0$ on $(0,1)$, and $H:=\int_0^1 h(z)\,dz>0$.

Define
$g^\varepsilon:\Theta\to\mathbb{R}$ by
$$
g^\varepsilon(\theta)=\chi(s(\theta))\left[h\!\left(\frac{\ell(\theta)-c(s(\theta))}{\gamma}\right)+h\!\left(\frac{\ell(\theta)+c(s(\theta))}{\gamma}\right)\right].
$$
$g^\varepsilon$ is smooth because $s$, $\ell$, $c$, $\chi$, and $h$ are. The
support of $g^\varepsilon$ is contained in the
union of the two sets
$$
S_\pm=\{\theta\in\Theta:|s(\theta)|\le\bar s,\ |\ell(\theta)\mp c(s(\theta))|\le\gamma\}.
$$
The strips $S_\pm$ have half-width $\gamma$ in the $\ell$-direction around $\ell=\pm c(s)$ (\Cref{fig:transport}). $S_+$ and $S_-$ are disjoint for all sufficiently small $\varepsilon$ because
$\gamma=\varepsilon^{5/4}$ and $c(s)=2\varepsilon \phi(s)\geq 2\varepsilon/B$. Moreover, $S_+$ and $S_-$ are contained in
$R$ for all sufficiently small $\varepsilon$ because, for $\theta\in S_\pm$,
$|\ell(\theta)|\le c(s(\theta))+\gamma=2\varepsilon \phi(s(\theta))+\gamma\leq 2\varepsilon B+\gamma$. Since $S_+\cap S_-=\emptyset$,
$0\le\chi\le1$, and $\|h\|_\infty\le1$, we have $\|g^\varepsilon\|_\infty\le1$.

For \eqref{eq:cap-mass}, integrate in the coordinates $(s,\ell)=T(\theta)$,
with $d\theta=\|q\|^{-1}\,ds\,d\ell$. For every fixed $s$, both integrals
$\int h\bigl((\ell\pm c(s))/\gamma\bigr)\,d\ell$ vanish because
$h$ is odd; hence $\int_\Theta g^\varepsilon(\theta)\,d\theta=0$ by Fubini.

\paragraph{Step 5: bounding the performance of $M^\varepsilon$ under $g^\varepsilon$.}

Fix $s\in[-\bar s,\bar s]$ with $\chi(s)>0$. Call $S_\pm^s:=\{\ell:|\ell\mp c(s)|\le\gamma\}$ the slice of $S_\pm$ at $s$. Since $S_\pm\subset R$, the payment on $S^s_+$ is $p_s$ for $\ell<c(s)$
and $p$ for $\ell>c(s)$, as noted in Step 2.
Similarly, the payment on $S^s_-$ is $p_s$ for $\ell>-c(s)$
and $0$ for $\ell<-c(s)$.
Integrating along the slice $S^s_+$
and substituting $\ell=c(s)+\gamma z$ yields
$$
\int h\!\left(\frac{\ell-c(s)}{\gamma}\right)t_{M^\varepsilon}(T^{-1}(s,\ell))\,d\ell
=\gamma\left[p_s\int_{-1}^0 h(z)\,dz+p\int_0^1 h(z)\,dz\right]
=(p-p_s)H\gamma.
$$
Similarly, integrating along the slice $S^s_-$ yields $(p_s-0)H\gamma$. The integrals along the two slices at $s$ sum to $pH\gamma$. By Fubini, integrating
over $s$ and applying the change of variables $d\theta=\|q\|^{-1}\,ds\,d\ell$ yields
$\REV(M^\varepsilon;g^\varepsilon)=A\gamma$ with $A=\frac{pH}{\|q\|}\int\chi(s)\,ds>0$,
which is the first half of \eqref{eq:cap-competitor}.

\paragraph{Step 6: bounding the performance of any menu with at most $k$ items under $g^\varepsilon$.}
Let $M'\in\mathcal{M}^{\le k}$. The types for which two given items in $\menu(M')$
yield equal utility form a line. Hence there is a set $\mathcal{L}$ of
at most $\binom{k}{2}$ lines such that, on each connected component of
$\Theta\setminus\bigcup\mathcal{L}$, the payment
$t_{M'}(\theta)$ is constant. Since $T$ is affine, $T(L)$ is a line in the
$(s,\ell)$-plane for every $L\in\mathcal{L}$. By individual rationality and by the normalization
$\Theta=[0,1]^2$, $t_{M'}(\theta)\in[0,2]$.

For $L\in\mathcal{L}$, let $E_\pm(L)=s(L\cap S_\pm)$ be the set of positions $s$ at
which $L$ meets $S_\pm$. If $s\in[-\bar s,\bar s]$ does not belong to $\bigcup_{L\in\mathcal{L}} E_+(L)$, then
$t_{M'}(T^{-1}(s,\ell))$ is constant in $\ell$ for $|\ell-c(s)|\le\gamma$, so the slice integral $\int h\left(\frac{\ell-c(s)}{\gamma}\right)t_{M'}(T^{-1}(s,\ell))\,d\ell$
vanishes because $h$ is odd.
The analogous statement holds for $S_-$. If $s\in[-\bar s,\bar s]$ belongs to $\bigcup_{L\in\mathcal{L}} E_+(L)$, then the slice integral has absolute
value at most $4\gamma H$ since $t_{M'}(\theta)\in[0,2]$ and $\int\left|h\left(\frac{\ell-c(s)}{\gamma}\right)\right|d\ell=2\gamma H$. The analogous statement holds for $S_-$.
Since $0\le\chi\le1$, integrating over $s$ as in Step 5 yields
$$
\lvert\REV(M';g^\varepsilon)\rvert\le\frac{4\gamma H}{\|q\|}\sum_{L\in\mathcal{L}}\bigl(\Leb(E_+(L))+\Leb(E_-(L))\bigr),
$$
so it remains to bound the Lebesgue measures $\Leb(E_\pm(L))$ of the sets $E_\pm(L)\subset\mathbb{R}$.

Fix an arbitrary $L\in\mathcal{L}$. If $T(L)=\{(s,\ell):s=s_0\}$ for some $s_0$, then
$E_\pm(L)\subseteq\{s_0\}$ has measure zero. Otherwise, $T(L)$ is the graph of
$\ell=\alpha s+\beta$ for some $\alpha,\beta\in\mathbb{R}$. For $s\in[-\bar s,\bar s]$, we then have $s\in E_\pm(L)$ if and only if $|d(s)|\le\gamma$
for $d(s)=c(s)\mp(\alpha s+\beta)$. Note
$d''(s)=2\varepsilon\phi''(s)>2\varepsilon/B$.
If $s\le s'$ lie in $E_\pm(L)$ and $d'(s)\ge0$ (so $d'(s')\ge0$ since $d''>0$), then integrating $d''(s)>2\varepsilon/B$ twice gives
$$
2\gamma\ge d(s')-d(s)\ge\varepsilon(s'-s)^2/B,
$$
so $E_\pm(L)\cap\{d'\ge0\}$ has diameter at most $\sqrt{2B\gamma/\varepsilon}$.\footnote{Concretely, for $\sigma\ge s$, $d'(\sigma)=d'(s)+\int_s^\sigma d''\ge\frac{2\varepsilon}{B}(\sigma-s)$ because $d'(s)\ge0$ and $d''>2\varepsilon/B$. Thus, $d(s')-d(s)=\int_s^{s'}d'\ge\frac{2\varepsilon}{B}\int_s^{s'}(\sigma-s)\,d\sigma=\frac{\varepsilon}{B}(s'-s)^2$. The upper bound comes from $|d(s)|,|d(s')|\le\gamma$.} Symmetrically, we obtain the same bound on the diameter of $E_\pm(L)\cap\{d'\le0\}$. Since these two sets cover $E_\pm(L)$, $\Leb(E_\pm(L))\le\sqrt{8B\gamma/\varepsilon}$. We conclude
$$
\lvert\REV(M';g^\varepsilon)\rvert\le\frac{4\gamma H}{\|q\|}\binom{k}{2}\cdot2\cdot\sqrt{8B\gamma/\varepsilon}=C_k\gamma^{3/2}\varepsilon^{-1/2}
$$
with $C_k=\frac{8\sqrt{8B}H}{\|q\|}\binom{k}{2}>0$.
Taking the supremum over $M'\in\mathcal{M}^{\le k}$ gives the second half of
\eqref{eq:cap-competitor}.

\paragraph{Step 7: wrapping up.}
Steps 2--6 establish \eqref{eq:cap-mass}--\eqref{eq:cap-competitor} for all
sufficiently small $\varepsilon$, which defines $\varepsilon_0$. For
\eqref{eq:cap-conclusion}, choose $\bar\varepsilon\in(0,\varepsilon_0]$ such
that $\sqrt{\bar\varepsilon}<\min f$ and
$A-C_k\bar\varepsilon^{1/8}-D\bar\varepsilon^{1/4}>0$, and fix
$\varepsilon\in(0,\bar\varepsilon)$. By \eqref{eq:cap-mass},
$f^\varepsilon=f+\sqrt{\varepsilon}g^\varepsilon$ is a probability density in
$C^\infty(\Theta)$ with $\|f^\varepsilon-f\|_\infty\le\sqrt{\varepsilon}$. For
every $M'\in\mathcal{M}^{\le k}$, we have $\REV(M';f)\le\REV^*(f)$, so by
\eqref{eq:cap-loss} and \eqref{eq:cap-competitor} and
$\gamma=\varepsilon^{5/4}$,
$$
\begin{aligned}
\REV(M^\varepsilon;f^\varepsilon)-\REV(M';f^\varepsilon)
&\ge-D\varepsilon^2+\sqrt{\varepsilon}\bigl(A\varepsilon^{5/4}-C_k\varepsilon^{11/8}\bigr)\\
&=\varepsilon^{7/4}\bigl(A-C_k\varepsilon^{1/8}-D\varepsilon^{1/4}\bigr)>0.
\end{aligned}
$$
Taking the supremum over $M'\in\mathcal{M}^{\le k}$ yields
$\REV^*(f^\varepsilon)\ge\REV(M^\varepsilon;f^\varepsilon)>\REV^{\le k}(f^\varepsilon)$.
\end{proof}

\begin{proof}[Proof of \Cref{thm:master}: Denseness]
    The smooth, strictly positive densities are dense in $\mathcal{F}$ by the Stone--Weierstrass Theorem; hence $\mathcal{F}^{>k}$ is dense in $\mathcal{F}$ for all $k\geq2$ by \Cref{lemma:cap}, and hence for all $k\in\mathbb{N}$ because $\mathcal{F}^{>k}\supseteq\mathcal{F}^{>2}$ for $k<2$. Continuous densities are $L^1$-dense in the set of all probability densities on $\Theta$, and Lebesgue absolutely continuous probability measures are weakly dense in the set of all probability measures on $\Theta$. Hence, $\bar{\mathcal{F}}^{>k}$ is dense in $\bar{\mathcal{F}}$, and $\mathcal{P}^{>k}$ is dense in $\mathcal{P}$.
\end{proof}

\subsection{Uniqueness}
\label{subsec:uniqueness}
\begin{proof}[Proof of \Cref{cor:uniqueness}]
    We identify each mechanism $M=(x,t)\in\mathcal{M}$ with an element of $L^1(\Theta;\mathbb{R}^3)$ so that uniqueness of the optimal mechanism up to null sets of types is uniqueness in $\mathcal{M}$. Equivalently, we identify IC and IR mechanisms with the associated indirect buyer utility function (\citet[Lemma~3]{lahr2024extreme}).  In particular, if the optimal mechanism is unique under this identification, then the optimal seller-favorable mechanism is unique.\footnote{Uniqueness holds by definition. For existence and, in particular, Borel-measurability, see Appendix A.1 in \citet{hart2015maximal}.}
    The set $\mathcal{M}$ is $L^1$-compact (\citet[Lemmas~3 and~13]{lahr2024extreme}), hence so is the projection $\mathcal{T}=\{t_M:M\in\mathcal{M}\}$. Let $\mathcal{T}^*(f)=\{t_M:M\in\mathcal{M}^*(f)\}$ and
    $
    \bar{\mathcal{F}}^*=\{f\in\bar{\mathcal{F}}:\mathcal{M}^*(f)\text{ is a singleton}\}.
    $

    We show that $\bar{\mathcal{F}}^*$ is a $G_\delta$ in $\bar{\mathcal{F}}$ and that $\bar{\mathcal{F}}^*\cap\mathcal{F}$ is dense in $\mathcal{F}$. Then, $\bar{\mathcal{F}}^*\cap\mathcal{F}$ is a dense $G_\delta$ in $\mathcal{F}$ and $\bar{\mathcal{F}}^*$ is a dense $G_\delta$ in $\bar{\mathcal{F}}$ because $\mathcal{F}$ is $L^1$-dense in $\bar{\mathcal{F}}$ and equipped with a stronger topology than $\bar{\mathcal{F}}$. Intersecting with the dense $G_\delta$ from \Cref{thm:master} in each space and applying the Baire Category Theorem completes the proof.

    Note $ \bar{\mathcal{F}}^*=\bigcap_{n\in\mathbb{N}}\bar{\mathcal{F}}_n^*$ with $ \bar{\mathcal{F}}_n^*=\{f\in\bar{\mathcal{F}}:\operatorname{diam}\mathcal{M}^*(f)<1/n\}$; hence  $ \bar{\mathcal{F}}^*$ is a $G_\delta$ if $ \bar{\mathcal{F}}_n^*$ is open for all $n\in\mathbb{N}$. The latter indeed follows from Berge's Maximum Theorem because
    $(M,f)\mapsto\int t_Mf$ is continuous on $\mathcal{M}\times L^1(\Theta)$ since $0\leq t_M\leq 2$ by IR.

    For denseness, define the function
    $
    h(f)=\max_{t\in\mathcal{T}}\int t(1+f)
    $
    on the separable Banach space $C_0=\{f\in C(\Theta):\int f=0\}$. The maximum is attained because $\mathcal{T}$ is $L^1$-compact and $t\mapsto\int t(1+f)$ is $L^1$-continuous for $f\in C(\Theta)$. The function $h$ is convex because it is the maximum of a family of affine functionals and Lipschitz continuous because $0\leq t\leq2$ for all $t\in\mathcal{T}$. By Theorem 4.17 in \citet{benyamini2000geometric}, the set $G\subseteq C_0$ on which $h$ is G\^{a}teaux differentiable is dense in $C_0$. In particular, $h$ has unique subgradients on $G$. If $t\in\mathcal{T}$ attains the maximum in the definition of $h$ at $f\in C_0$, i.e., $t\in\mathcal{T}^*(1+f)$, then the linear functional $g\mapsto\int tg$ is a subgradient of $h$ at $f$. Hence, if $f\in G$ and $t,t'\in\mathcal{T}^*(1+f)$, $t-t'$ is almost everywhere constant. But $0\leq t(\theta),t'(\theta)\leq \theta_1+\theta_2$ by IR, so this constant is zero. Hence $\mathcal{T}^*(1+f)$ is a singleton for all $f\in G$.

    If $f\in\mathcal{F}$ is strictly positive, then every $t\in\mathcal{T}^*(f)$ is undominated in that there is no $t'\in\mathcal{T}$ with $t'\geq t$ almost everywhere and $t'>t$ on a set of positive measure. By Lemma 9 in \citet{lahr2024extreme}, if $t\in\mathcal{T}$ is undominated, then there exists a unique $M\in\mathcal{M}$ with $t_M=t$. Thus, if $f\in (1+G)\cap\mathcal{F}$ is strictly positive, then $f\in\bar{\mathcal{F}}^*$. Consequently, $\bar{\mathcal{F}}^*\cap \mathcal{F}$ is dense in $\mathcal{F}$ because the strictly positive densities in $\mathcal{F}$ form an open set in $(1+C_0)$ and because $G$ is dense in $C_0$.
\end{proof}

\subsection{Frictions}
\label{subsec:frictions}

\begin{proof}[Proof of \Cref{lemma:menu_costs}]
For $\lambda>0$, only finitely many menu sizes yield non-negative profit, since revenue is bounded and costs are unbounded. \Cref{lemma:bounded-attainment} therefore yields an optimal mechanism. Since $\REV^{\leq k}(\mu)-\lambda C(k)$ has decreasing differences in $(k,\lambda)$, its smallest maximizer $k_\lambda(\mu)$ is non-increasing in $\lambda$ by Topkis' theorem, with $k_\lambda(\mu)\leq k_0(\mu)$.

We have $\REV^{\leq k}(\mu)\to\REV^*(\mu)$ as $k\to\infty$ (\citet[Proposition~9.2]{hart2019selling}). Thus, for every $k<k_0(\mu)$, \Cref{lemma:bounded-attainment} gives $k'>k$ with $\REV^{\leq k'}(\mu)>\REV^{\leq k}(\mu)$. A mechanism attaining $\REV^{\leq k'}(\mu)$ has profit converging to $\REV^{\leq k'}(\mu)$ as $\lambda\to0$. But every mechanism with menu size at most $k$ yields profit at most $\REV^{\leq k}(\mu)$. Thus, $k_\lambda(\mu)>k$ for all sufficiently small $\lambda>0$. Since $k<k_0(\mu)$ was arbitrary and $k_\lambda(\mu)\leq k_0(\mu)$ for all $\lambda\geq 0$, $k_\lambda(\mu)\to k_0(\mu)$.
\end{proof}

\begin{proof}[Proof of \Cref{cor:discrete}]
Fix any $k \in \mathbb{N}$. Recall that $\mathcal{P}^{>k}$ is the set of type
distributions for which all optimal mechanisms have menu size $>k$, and let
$\mathcal{P}^{\leq k}=\mathcal{P}\setminus\mathcal{P}^{>k}$. For the rest of
the proof, let $\mathcal{P}^{>k}(Q)$ and $\mathcal{P}^{\leq k}(Q)$ be the same
objects for the model where the allocation space is $Q\subseteq [0,1]^2$ instead of the unit square. Since the dense $G_\delta$'s in \Cref{thm:master}
are $\bigcap_{k\in\mathbb{N}} \mathcal{P}^{>k}$, $\bigcap_{k\in\mathbb{N}} \bar{\mathcal{F}}^{>k}$, and
$\bigcap_{k\in\mathbb{N}} \mathcal{F}^{>k}$, and since $\mathcal{F}^{>k} \subseteq
\bar{\mathcal{F}}^{>k} \subseteq \mathcal{P}^{>k}$, it suffices to prove the
following claim.

We claim that for all $\mu \in \mathcal{P}^{>k}$ there exists $\delta_\mu > 0$
such that if $\delta:=\mesh(Q) < \delta_\mu$, then $\mu \in \mathcal{P}^{>k}(Q)$. This can be shown with a ``scale-and-round'' argument (akin to, e.g., \citet[Lemma~4.2]{dughmi2014sampling} or \citet[Lemma~9.3]{hart2019selling}). Take any optimal
mechanism $M=(x,t)$ for $\mu$. Let $\tilde t(\theta)=(1-\eta)t(\theta)$ for
some $\eta\in (0,1)$. Let $\tilde x(\theta)$ be one of the
nearest allocations to $x(\theta)$ in $Q$ with respect to the 1-norm $\|\cdot\|_1$. $\tilde M = (\tilde x,\tilde t)$ is generally not IC, but rational choice from the closure of
$\menu(\tilde M)$ yields an IC and IR mechanism. If $\theta$
chooses $(\tilde x(\theta);\tilde t(\theta))$, then the revenue loss is $\eta
t(\theta)$. If $\theta$ chooses $(\tilde x(\theta');\tilde t(\theta'))$  (or some limit point thereof), then
$$
\tilde t(\theta')-\tilde t(\theta)
\leq(\tilde x(\theta')-\tilde x(\theta))\cdot\theta
\leq(x(\theta')-x(\theta))\cdot\theta+2\delta
\leq t(\theta')-t(\theta)+2\delta.
$$
The first inequality follows from rational choice from $(\tilde x,\tilde t)$, the second inequality from $\delta=\mesh(Q)$ and the definition of $\tilde x$, and the third inequality from IC of $(x,t)$.
Rearranging yields $\tilde t(\theta')\geq (1-\eta)t(\theta)-2\delta\frac{1-\eta}{\eta}$. We conclude that
the revenue loss vanishes pointwise as $\eta\to 0$ and $\delta/\eta\to 0$. Since $\mu\in \mathcal{P}^{>k}$, $\REV(M;\mu)-\REV^{\le k}(\mu)>0$ by \Cref{lemma:bounded-attainment}. We conclude that $\mu\notin\mathcal{P}^{\le k}(Q)$ if $\delta=\mesh(Q)$ is sufficiently small.
\end{proof}

\begin{proof}[Proof of \Cref{lemma:attention}]
Fix $\mu\in\mathcal{P}$. For a mechanism $M$ with menu size $k$, let $u_M(\theta)=x_M(\theta)\cdot\theta-t_M(\theta)$ be type $\theta$'s indirect utility.
A buyer with belief $\beta$ inspects the menu if and only if $\int_\Theta u_M\,d\beta\geq\lambda C(k)$ (breaking ties in favor of inspection).  Thus, for a given belief distribution $\tau$ averaging to $\mu$, $M$ yields revenue
$$
\REV_\lambda(M;\tau)=\int_{\mathcal{P}}\mathbf{1}\Bigl\{\int_\Theta u_M\,d\beta\geq\lambda C(k)\Bigr\}\int_\Theta t_M\,d\beta\,d\tau(\beta).
$$

We first argue that an optimal mechanism exists. Fix an arbitrary belief distribution $\tau$ averaging to $\mu$. For every $\lambda>0$ there exists a menu size $\bar k\in\mathbb{N}$ above which every mechanism yields zero revenue since we have assumed that $C$ is unbounded, as in \Cref{subsec:menucosts}, and since surplus is bounded because the type space is bounded.
Let $(M^n)_{n\in \mathbb{N}}\subset\mathcal{M}^{\leq \bar k}$ be a sequence of mechanisms with $\REV_\lambda (M^n;\tau)\to\sup_{M'\in\mathcal{M}} \REV_{\lambda} (M';\tau)$.
Since $|\menu(M^n)|\leq \bar k$ for all $n\in\mathbb{N}$, we may assume $|\menu(M^n)|=k\leq\bar k$ up to taking a subsequence.
Define the limit mechanism $M$ as in the proof of \Cref{lemma:bounded-attainment}. By construction, $|\menu(M)|\leq k$. By Proposition 5 in \citet{hart2026existence} applied to belief type $\beta$, $\limsup_n \int t_{M^n}\,d\beta\leq \int t_M\,d\beta$.
By construction, $u_{M^n}\to u_{M}$ uniformly. Hence, any belief type $\beta$ who inspects infinitely often along the sequence $(M^n)$ also inspects $M$ since $|\menu(M)|\leq k$ and $C$ is increasing. Thus,
$$
\limsup_n \mathbf{1}\Bigl\{\int_\Theta u_{M^n}\,d\beta\geq\lambda C(k)\Bigr\}\int t_{M^n}\,d\beta\leq \mathbf{1}\Bigl\{\int_\Theta u_M\,d\beta\geq\lambda C(|\menu(M)|)\Bigr\}\int t_M\,d\beta.
$$
Since the payment rules are bounded by IR, we conclude $\limsup_n \REV_\lambda (M^n;\tau)\leq \REV_\lambda (M;\tau)$ by the Reverse Fatou lemma, so $M$ attains the supremum.

We now argue convergence as $\lambda\to 0$. Fix any  $k<k_0(\mu)$ and choose $k'>k$ with $\REV^{\leq k'}(\mu)>\REV^{\leq k}(\mu)$. Take a mechanism $ \hat M\in\mathcal{M}^{\leq k'}$ attaining $\REV^{\leq k'}(\mu)$ (which exists by \Cref{lemma:bounded-attainment}). Obtain a mechanism $M^\varepsilon\in \mathcal{M}^{\leq k'}$ by multiplying the prices of $\hat M$ by $1-\varepsilon$, where $0<\varepsilon<1$, and then letting each buyer type choose their favorite option (which may be different from what they chose under $\hat M$). In $M^\varepsilon$, every type $\theta$'s preferred option yields utility at least $\varepsilon t_{\hat M}(\theta)$ and payment at least $(1-\varepsilon)t_{\hat M}(\theta)$, by IR and IC before and after the discount, respectively. Every belief type $\beta$ with $\int_\Theta t_{\hat M}\,d\beta\geq\lambda C(k')/\varepsilon$ therefore inspects $M^\varepsilon$.
Belief types that do not inspect $M^\varepsilon$ therefore imply a revenue loss of at most $\lambda C(k')/\varepsilon$ relative to $\REV(\hat M;\mu)$, so $\REV_\lambda(M^\varepsilon;\tau)\geq(1-\varepsilon)\bigl(\REV(\hat M;\mu)-\lambda C(k')/\varepsilon\bigr)$ uniformly over all belief distributions $\tau$ averaging to $\mu$. If we choose $\varepsilon$ small enough and then $\lambda$ small enough, then the right-hand side exceeds $\REV^{\leq k}(\mu)$ since $\REV(\hat M;\mu)>\REV^{\leq k}(\mu)$. Thus, no mechanism of menu size at most $k$ can be optimal, uniformly over all belief distributions $\tau$ averaging to $\mu$.
Since $k<k_0(\mu)$ was arbitrary, the desired conclusion follows.
\end{proof}

\section{Examples}\label{app:examples}

\begin{proof}[Calculations for \Cref{example:setup_costs}]
    Let the first group receive $(q;p)$ and the second $(q';p')$. By \eqref{eq:IR} for the first group and \eqref{eq:IC} for the second, $p\leq\tfrac45q_1+\tfrac15q_2$ and $p'\leq q'_1+q'_2-q_1-q_2+p$, so $\REV(M;\mu)=\tfrac{7}{10}p+\tfrac{3}{10}p'\leq\tfrac12q_1-\tfrac1{10}q_2+\tfrac3{10}(q'_1+q'_2)\leq\tfrac{11}{10}$, with equality only if $(q;p)=(1,0;\tfrac45)$ and $(q';p')=(1,1;\tfrac95)$. If a menu contains only one option besides the outside option, then revenue is at most $1$ if the first group buys and $\tfrac{3}{10}\cdot2$ otherwise. Thus, the best menus of sizes two and three yield profits $1-\lambda$ and $\tfrac{11}{10}-2\lambda$, respectively, giving the cutoff $\lambda=\tfrac{1}{10}$. Consumer surplus and total surplus are easily calculated.
\end{proof}

\begin{proof}[Calculations for \Cref{example:cognitive_costs}]
   Let us label the three groups of buyers $A$, $B$, and $L$, with types $\theta_A=(1,0)$, $\theta_B=(0,\tfrac{1}{2})$, and $\theta_L=(\tfrac{1}{8},\tfrac{1}{8})$ (in proportions $(\tfrac{1}{10},\tfrac{2}{10},\tfrac{7}{10})$).
Fix any menu, and let $k$ denote the size.
Let $(x_1,x_2)$ and $t_L$ be the equilibrium allocation and payment, respectively, of buyer $L$. Let $t_A$ and $t_B$ be the equilibrium payments of buyers $A$ and $B$, respectively.

First, consider menus that $L$ will inspect. Then, $t_L\leq \tfrac{x_1+x_2}{8}-\lambda C(k)$ (where, recall, $C(k)=k-1$).
If $A$ and $B$ also inspect, then they must prefer their respective choices to that of $L$. Thus, we get $1-t_A\geq x_1-t_L$ and $\tfrac12-t_B\geq \tfrac12x_2-t_L$. If $A$ or $B$ does not inspect, then these inequalities hold trivially because their payment is zero. Combining these observations, revenue is at most
$$
\tfrac{1}{10}t_A+\tfrac{2}{10}t_B+\tfrac{7}{10}t_L\leq \tfrac15-\tfrac{1}{10}(x_1+x_2)+t_L\leq \tfrac14-\lambda C(k).
$$
Offering the grand bundle $(1,1)$ at price $\tfrac{1}{4}-\lambda$ attains this bound. This mechanism can be optimal only if $\lambda\leq \tfrac{1}{4}$.

Now, consider menus that $L$ will not inspect.
Among menus of size 3, it is optimal to offer $(1,0)$ and $(0,1)$ at prices $t_A=1-2\lambda$ and $t_B=\tfrac{1}{2}-2\lambda$ because these extract the full surplus, yielding revenue
$$
\tfrac1{10}(1-2\lambda)+\tfrac2{10}(\tfrac{1}{2}-2\lambda)=\tfrac2{10}(1-3\lambda)
$$
for $\lambda\leq \tfrac{1}{4}$. For $\lambda>\tfrac{1}{4}$, $B$ does not inspect, so it is better to offer a menu of size 2.
Among menus of size 2, it is optimal to offer the grand bundle because it maximizes surplus, either at price $\tfrac{1}{2}-\lambda$ ($A$ and $B$ buy) or price $1-\lambda$ (only $A$ buys). This yields revenue $$\max\{\tfrac3{10}(\tfrac{1}{2}-\lambda),\tfrac1{10}(1-\lambda)\}=\tfrac1{10}(1-\lambda)+\max\{0,\tfrac{1-4\lambda}{20}\}.$$
Thus, the latter is  better for $\lambda>\tfrac{1}{4}$ and the former for $\lambda<\tfrac{1}{4}$.

Putting these observations together, we see that $(1,1)$ at price $\tfrac{1}{4}-\lambda$ is optimal for $\lambda\in(0,\tfrac{1}{8})$, $(1,0)$ and $(0,1)$ at prices $1-2\lambda$ and $\tfrac{1}{2}-2\lambda$ are optimal for $\lambda\in(\tfrac{1}{8},\tfrac{1}{6})$, and the bundle $(1,1)$ at the optimally set price is optimal for $\lambda\in(\tfrac{1}{6},1)$.
\end{proof}

\printbibliography
\end{document}